\documentclass[11pt,onecolumn,letter]{IEEEtran}
\usepackage{etoolbox}
\patchcmd{\section}{\centering}{}{}{}

\usepackage{geometry}
\usepackage{graphicx}

\usepackage[cmex10]{amsmath} 
\usepackage{amssymb}
\usepackage{amsfonts}

\usepackage{cuted}

\usepackage{subcaption}

\usepackage[noadjust]{cite}
\usepackage{epstopdf}

\usepackage[x11names]{xcolor}  

\usepackage[disable,colorinlistoftodos]{todonotes}  

\usepackage[T1]{fontenc}
\usepackage{charter}
\usepackage{environ}
\usepackage{tikz}
\usetikzlibrary{calc,matrix}

\usepackage[section]{placeins} 
\usepackage{stackengine}

\usepackage{enumitem}
\setlist[enumerate]{label=\roman*} 

\usepackage{microtype}

\newtheorem{theorem}{Theorem}
\newtheorem{corollary}{Corollary}
\newtheorem{conjecture}{Conjecture}

\usepackage[colorlinks]{hyperref}
\hypersetup{%
	colorlinks = true,
	citecolor=Green4,
	linkcolor  = blue
}

\pagebreak

\ifCLASSINFOpdf
\else
\fi

\begin{document}
	\IEEEoverridecommandlockouts
	
	
	
	
	\title{Sensing Method for Two-Target Detection in Time-Constrained Vector Poisson Channel}
	\author{\IEEEauthorblockN{Muhammad Fahad\IEEEauthorrefmark{1},~\IEEEmembership{Member,~IEEE}, and Daniel R. Fuhrmann\IEEEauthorrefmark{2},~\IEEEmembership{Fellow,~IEEE}} \\
		\IEEEauthorblockA{Department of Applied Computing,\\
			Michigan Technological University\\ 
			Houghton, MI 49931, USA \\
			\IEEEauthorrefmark{1}mfahad@mtu.edu,
			\IEEEauthorrefmark{2}fuhrmann@mtu.edu}}
	\maketitle
	\begin{abstract}
		It is an experimental design problem in which there are two Poisson sources with two possible and known rates, and one counter.  Through a switch, the counter can observe the sources individually or the counts can be combined so that the counter observes the sum of the two.  The sensor scheduling problem is to determine an optimal proportion of the available time to be allocated toward individual and joint sensing, under a total time constraint.  Two different metrics are used for optimization: mutual information between the sources and the observed counts, and probability of detection for the associated source detection problem.  Our results, which are primarily computational, indicate similar but not identical results under the two cost functions.
	\end{abstract}
	\section{Introduction}  \label{intro}
	The importance of sensor scheduling emerges when it is not possible to collect unlimited amounts of data due to resource constraints. \cite{hero2007foundations}. Generally speaking it is always advantageous to collect as much data as possible, provided such data can be properly managed and processed.  However there are constraints imposed by available time, computational resources, memory requirements, communication bandwidth, available battery power, and sensor deployment patterns.  This creates a need for an optimal or heuristic scheduling methodology to extract meaningful data from different available data sources in an efficient manner.  
	
	One of the major challenges in sensor scheduling problems involving switching among various sources arises from the combinatorial nature of the possible switching possibilities.  This problem becomes further complicated if the sensor scheduling is done in continuous-time settings \cite{lee2001sensor}.
	\begin{figure}[t]
			\centering
			\includegraphics[width=.25\linewidth]{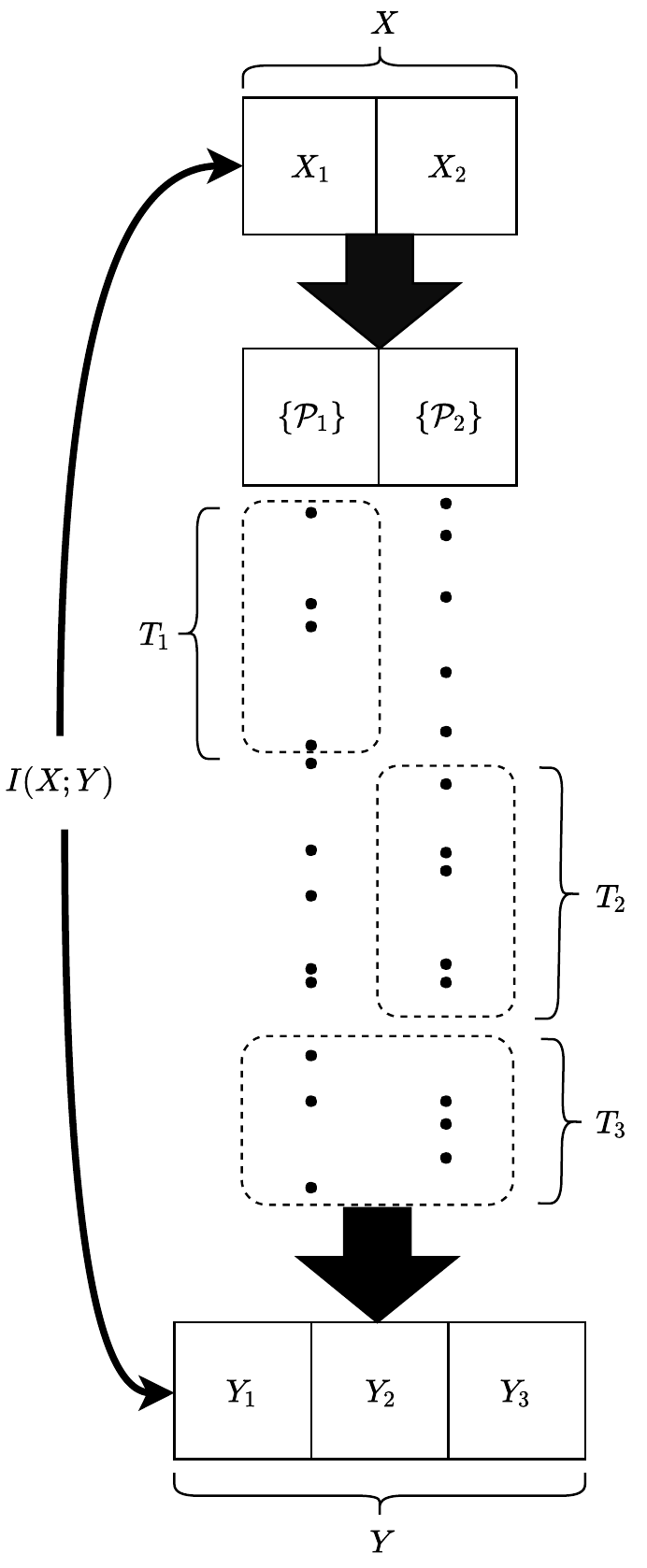}
			\caption{Illustration of sensing method for Bayesian detection of $2-$long hidden random vector $X$ from $3-$long observable random vector $Y$ in vector Poisson channel under a total observation time constraint, where $\{ \mathcal{P}_i \}$ is a conditional Poisson point process given Bernoulli distributed input $X_i$.  \todo[inline]{1) Open draw.io in chrome. \newline 2) Open file from C:$\backslash$Users$\backslash$mfahad$\backslash$Dropbox$\backslash$Apps$\backslash$drawio$\backslash$Sensing\_Paradigm2.html \newline 3) Export to HTML and save in the same folder.  \newline 4) Open HTML and print PDF \newline 5) Crop the pdf using acrobats tools: crop  \newline 6) Save the pdf in folder C:$\backslash$Users$\backslash$mfahad$\backslash$Dropbox$\backslash$PhD\_Paper\_1 }}
			\label{f1}
		\end{figure}
		\begin{figure*}[!h] 
			\centering
			\includegraphics[width=\textwidth,trim={0inch 0inch 0inch 0inch},clip=true]{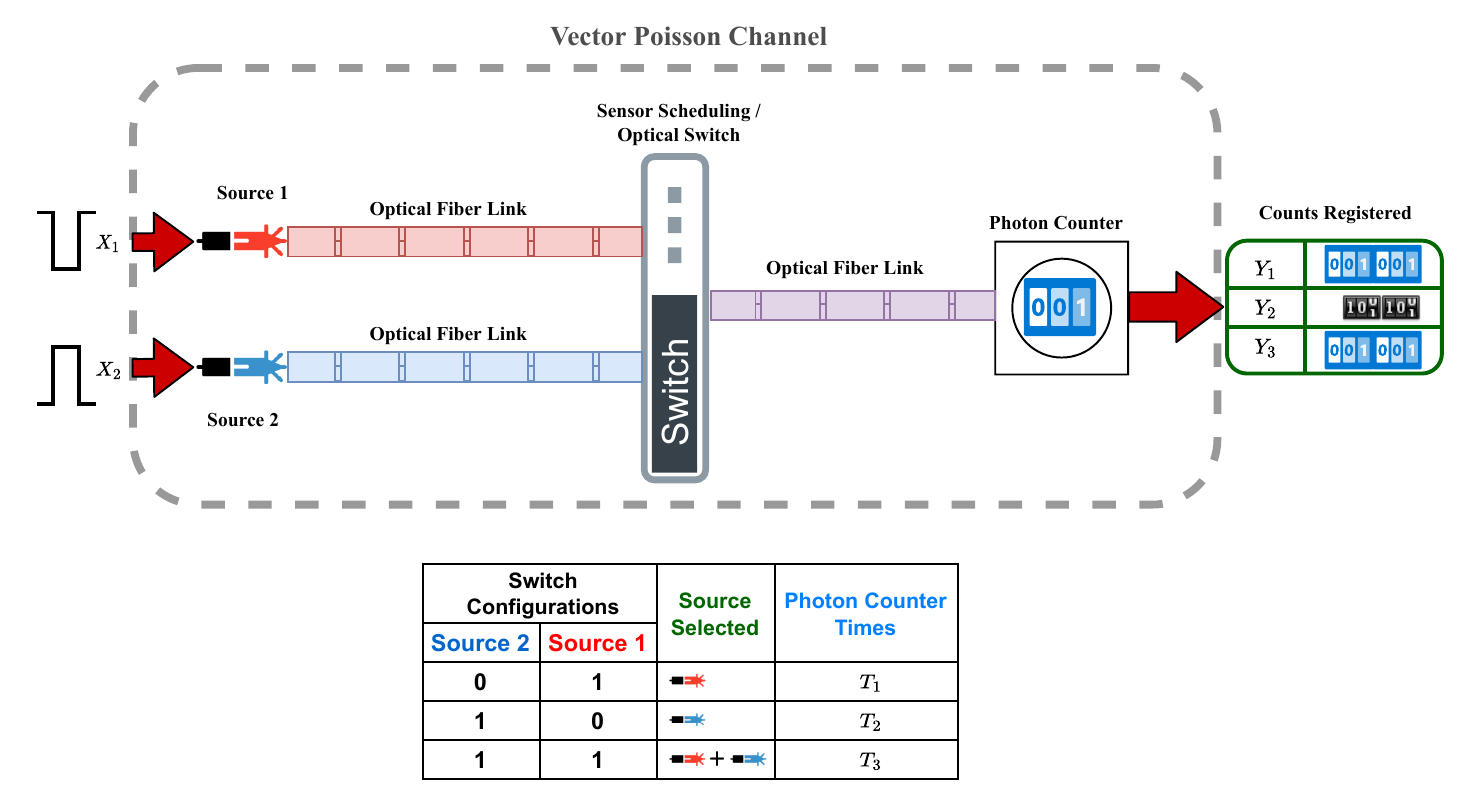}
			\caption{An abstract example: $Y_1$, $Y_2$ and $Y_3$ are total counts of photons accumulated in times $T_1$, $T_2$ and $T_3$, respectively. LED source 1 (and source 2) initiates a homogeneous Poisson process of intensity $\lambda_0$ if Bernoulli distributed random input signal $X_1$ (and $X_2$) takes value of $0$ with probability $(1-p)$; and $\lambda_1$ if input signal $X_1$ (and $X_2$) takes value of $1$ with probability $p$. Once the input random vector $[X_0 \: X_1]$ assumes any of the four possible states $[X_0 \: X_1: 00, 01, 10, 11]$, it doesn't change its state during the course of Photon counting for given time $T$.}
			\label{f21}
		\end{figure*}	
		
		Information-theoretic approaches have long been used in sensor scheduling problems, especially for their invariance to invertible transformations of the data. Many of the early approaches towards sensor management were information-theoretic, for instance \cite{manyika1992information}, \cite{schmaedeke1993information}.  Mutual information between observed data and unknown variables under test provides an intuitively appealing scalar metric for the performance of a given data acquisition methodology.  That being said, it is also well-known that maximizing mutual information does not necessarily lead to optimal detection or estimation performance.
		
		In this paper we address a very basic problem of \emph{scheduling} a single sensor with multiple switch configurations to detect a binary $ 2 $-long random vector in a vector Poisson channel.  This might be of special interest in context of Poisson \emph{compressive sensing}, where sparse signal conditions can be exploited for reduced computational cost and efficient sensor scheduling. We address this particular sensor scheduling problem from both \emph{information theoretic} and \emph{detection theoretic} perspectives.  As will be seen, this problem is deceptively simple yet does not readily admit closed-form solutions.
		
		The Poisson channel has been traditionally difficult to work with due to two inherent obstacles attached with it. First, added Poisson noise and scaling of input does not consolidate into a single parameter as SNR does in Gaussian channel. Second, scaling the support of a Poisson random variable (integers) does not result in another Poisson random variable. This is in contrast to Gaussian channels which are relatively well-studied from both information-theoretic and estimation-theoretic aspects \cite{yang2007mimo},  \cite{payaro2009hessian}, \cite{verdu2010mismatched}.  The Poisson channel has the advantage in thought experiments such as ours of simple conceptual models for data acquisition based on counting (photons, say) and switches that either allow counts to either pass through or be discarded.
		
		As will be seen when our problem is stated in more detail, fundamentally we are exploring a trade-off between SNR and ambiguity that arises when phenomena from multiple sources are combined in a single observation.  Adding multiple Poisson processes results in a Poisson process at a higher rate; higher rates correspond to higher SNR, and generally higher SNR is seen as a good thing for detection and estimation performance.  However, this SNR comes at a cost: with a single detector looking at counts from multiple processes, the observed Poisson counts cannot be disambiguated back to their original sources.  In our proposed scheme, part of the time is spent looking at sources individually (to help with this disambiguation) and part of the time is spent observing the sources jointly (to increase SNR).  Finding the right balance between these two approaches is the heart of our problem.
		
		In an unconstrained sensing-time setting, the vector Poisson channel has gained much research activity and interesting results are shown paralleling to that of vector Gaussian channel. One of the seminal works in linking the \emph{information theory} and \emph{estimation theory} in the context of scalar Poisson channel is done by Guo \textit{et al.} in \cite{guo2008mutual}, where the derivative of mutual information with respect to signal intensity is equated with a function of conditional expectation; providing a ground for the possible Poisson counterpart of a Gaussian channel. This result for scalar Poisson channel was further refined by Atar and Weissman in \cite{atar2012mutual} where exact relationship between derivative of mutual information and minimum mean loss error is given by providing the loss function $ l(x,\hat{x})=x \operatorname{log}(\frac{x}{\hat{x}})-x+\hat{x} $ which shares a key property with squared error loss i-e optimality of conditional expectation with respect to mean loss criterion. This result together with $	\frac{d}{d \alpha} I(X;\mathcal{P}(\alpha X)) \Big |_{\alpha=0}=E[X \cdot \operatorname{log}{X}]-E[X]\cdot \operatorname{log}(E[X]) $ given in \cite{guo2008mutual} provides us an exact answer to the question: for a given finite sensing time $T$ and prior $p$, which of the two sensing mechanisms, individual or joint sensing, is better than the other? (however, hybrid sensing still remains elusive and this is we have investigated in this paper). 
		
		Wang \textit{et al.} \cite{wang2014bregman} unifies the vector Poisson and Gaussian channels by constructing a generalization of the classical Bregman divergence and extended the scalar result to vector case (unconstrained). They provide the gradient of mutual information with respect to their input scaling matrices for both Poisson and Gaussian channel. But, for a general vector Poisson channel existence of gradient of mutual information w.r.t scaling matrix is defined in terms of expected value of the Bregman divergence matrix with a strictly K-convex loss function and which requires the partial ordering interpretation \cite{boyd2004convex} \cite{wang2014bregman} and which doesn't unify our problem. We are therefore also interested in exploring the general concave nature of our problem w.r.t scaling matrix (if it exists), which is discussed in coming sections.
		
		The rest of the paper is organized as follows: Section \ref{probdes} introduces the problem description, explaining the Poisson vector channel in which the target is to be observed. Section \ref{detofc} provides the \emph{information theoretic} model, while Section \ref{detd} describes the \emph{detection theoretic} model of the problem. Section \ref{comand} discusses the computed results from the previous two sections. Finally, Section \ref{con} concludes the paper.	
		\textit{Notation:} Upper case letters denote random vectors. Realizations of the random vectors are denoted by lower case letters. A number in subscript is used to show the component number of the random vector. We use $ X_1 $ and $ Y_1 $ to represent scalar input and output random variables, respectively. The superscript $ (\cdot)^{\intercal}  $ denotes the matrix/vector transpose. Deterministic time variable $ T_i $ is used to denote the sensing time for $ i^{th} $ conditional Poisson process and $ T $ is a given finite time. $ \alpha $ is an arbitrary positive scalar variable. $ \Phi $ represents the scaling matrix. $ p $ is the prior probability of 1 for a Bernoulli random variable.  $ f_X(x) $ denotes the probability mass function of $ X $, and the subscript will normally be dropped to simplify notation.  $\operatorname{Poiss}(y;\lambda ) \equiv \frac{e^{-\lambda }\lambda ^{y}}{y!}$.
		\begin{figure}[t!] 
			\centering
			\includegraphics[width=0.7\linewidth]{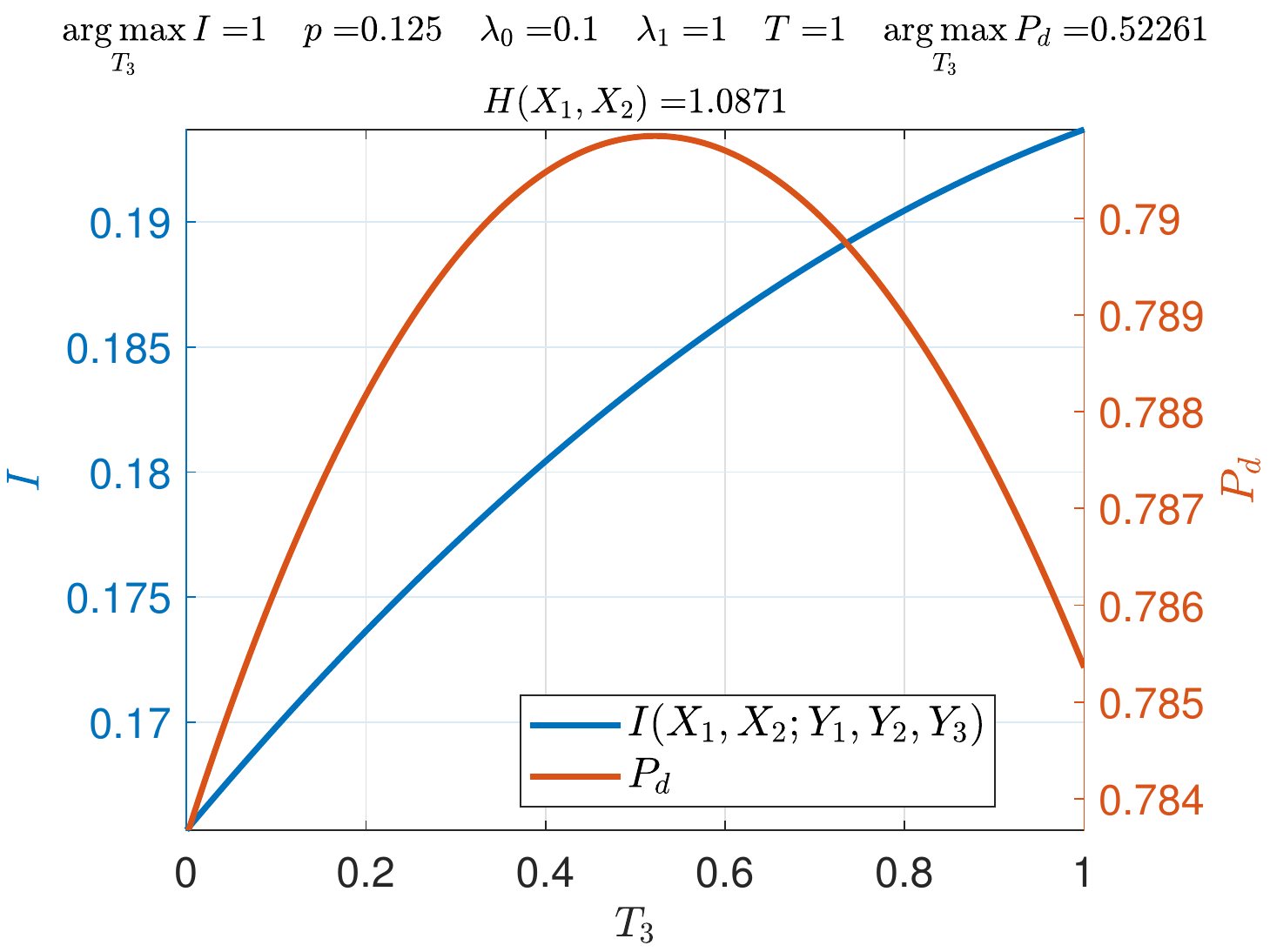}  
			\caption{Mutual information $ I(X;Y) $ and Bayesian probability of total correct detections $P_d$ vs. time $T_3$ where $0 \le T_3 \le T$  and $T_1=T_2=\frac{T-T_3}{2}$ such that time constraint $T=T_1+T_2+T_3$ is satisfied. \todo[disable,inline]{\texttt{\detokenize{Pd_Cd_MI_3.m}}                     }	}
			\label{f0}
		\end{figure}
		\section{Problem Description} \label{probdes}
		Let $X_1$ and $X_2$ be two independent and identical distributed (i.i.d) Bernoulli random variables with $ p $ being the probability of occurrence of $ 1 $. We may define probability mass function $ f $ of discrete random vector $ X \equiv [X_1,X_2]^{\intercal}$ as
		\begin{IEEEeqnarray*}{rCl}
			f_{X}(x)=\left\{
			\begin{array}{ll}
				p^2 & \quad x  =  \: [1 \quad1]^{\intercal}  \\ 
				(1-p)^2 & \quad x  =  \: [0 \quad0]^{\intercal} \\
				p(1-p) & \quad  x  =  \: [0\quad 1]^{\intercal} \quad \text{or} \quad  x  =  \: [1 \quad 0]^{\intercal}  \\             
			\end{array}
			\right. \label{w8} \IEEEyesnumber
		\end{IEEEeqnarray*}
		Let $\{ \mathcal{P}_1(t), t \ge 0 \}$ and $\{ \mathcal{P}_2(t), t \ge 0 \}$be two \emph{conditional point Poisson processes} \cite[pp. 88-89]{ross1996stochastic} such that their rate parameters, either $\lambda_0$ or $\lambda_1$ are determined by $X_1$ and $ X_2 $, respectively.  That is, $\lambda(x_i) = \lambda_0$ if $x_i = 0$ and $\lambda_1$ if $x_i = 1$, for $i=1,2$.  $\lambda_0$ and $\lambda_1$ are assumed known.
		We may count the number of arrivals from the two \emph{conditional Poisson arrival processes} in three possible configurations: two by individually counting the arrivals from the two processes $\mathcal{P}_1$, $ \mathcal{P}_2$ and one by counting the arrivals from the sum of two processes $  \{\mathcal{P}_1(t)+\mathcal{P}_2(t), t \ge 0 \} $ with given rate parameter $\lambda(X_1)+ \lambda(X_2)  $ as illustrated in Figs.(\ref{f1}) and (\ref{f21}).  The counting is performed such that at any given time, only one of the three possible configurations is active. Furthermore, because of the \emph{independent increments} property of \emph{conditional Poisson processes}, it is not necessary to switch back and forth among possible configurations; it is sufficient to be in configuration 1 for time $ T_1 $, followed by configuration 2 for time $ T_2 $, and then configuration 3 for time $ T_3 $. Additionally, counting is performed with a finite time constraint $ T=T_1+T_2+T_3 $ where $ T_1 $, $ T_2 $ and $ T_3 $ are the unknown time proportions in counting arrivals from processes $\{ \mathcal{P}_1(t), t \ge 0 \}$, $\{ \mathcal{P}_2(t), t \ge 0 \}$ and $  \{\mathcal{P}_1(t)+\mathcal{P}_2(t), t \ge 0 \} $, respectively and $ T $ is total time. After utilizing available time $ T $, the above counting paradigm leads to a multivariate Poisson mixture model with four component in three dimensions; we write random vector $ Y^{\intercal}\equiv[Y_1 \: \: Y_2 \: \: Y_3] $, so that $ Y \in \{ 0,1,2,3 \cdots \} ^3 $. Each $ Y_i $ is Poisson random variable given $ X $; such that their conditional law is,
		\begin{IEEEeqnarray*}{rCl}
			Y_1|X_1 & \:  \sim \: & \operatorname{Poiss}\Big(y_1; \big(\lambda_0 \cdot (1-X_1)+\lambda_1  \cdot X_1\big) \cdot T_1\Big)  \\
			Y_2|X_2 & \:  \sim \: & \operatorname{Poiss}\Big(y_2; \big(\lambda_0  \cdot (1-X_2)+\lambda_1 \cdot  X_2\big) \cdot T_2\Big)  \\
			Y_3|(X_1+X_2) & \: \sim \: & \operatorname{Poiss}\Big(y_3; \big((2-(X_1+X_2)) \cdot  \lambda_0 + \\ \> && (X_1+X_2) \cdot \lambda_1\big) \cdot T_3 \Big). \label{a} \IEEEyesnumber
		\end{IEEEeqnarray*} The observed random vector $ Y $ carries \emph{information} about hidden random vector $ X^{\intercal} \equiv [X_1 \: X_2]$.  We are interested in strategies for selecting the times $T_1$, $T_2$, $T_3$ that optimize some cost related to the inference problem, either mutual information or probability of correct detection.  Strategies that involve setting $T_3$ to 0 are referred to as \emph{individual sensing}; any time spent looking at combined counts is termed \emph{joint sensing}.  The general case, which involves some individual and some joint sensing, will be referred to as \emph{hybrid sensing}.  
		
		In matrix form we may relate the intensities of conditional Poisson distributed $ Y_i $'s as, 
		%

		\begin{equation}
			\label{e0} 
			\begin{aligned}
				\overbrace{	\begin{bmatrix}
						\mu_1  \\ \mu_2 \\ \mu_3
				\end{bmatrix}}^{\mathcal{M}: \: 3\times 1 }
				&=
				\overbrace{	\begin{bmatrix}
						T_1 & 0 & 0 \\
						0 & T_2 & 0 \\
						0 & 0 & T_3 \\
				\end{bmatrix}} ^{\mathcal{D}: \: 3\times 3 }
				\overbrace{	\begin{bmatrix}
						1 & 0 \\
						0& 1 \\
						1&  1
				\end{bmatrix}} ^{\mathcal{B}: \: 3 \times 2}
				\overbrace{			\begin{bmatrix}
						\Lambda_1   \\
						\Lambda_2   \\
				\end{bmatrix}} ^{\mathit{\Lambda}: \: 2 \times 1}\\
				&=
				\overbrace{	\begin{bmatrix}
						T_1 & 0  \\
						0 & T_2   \\
						T_3 & T_3 \\
				\end{bmatrix}} ^{\Phi: \: 3\times 2 }
				\overbrace{			\begin{bmatrix}
						\Lambda_1   \\
						\Lambda_2   \\
				\end{bmatrix}} ^{\mathit{\Lambda}: \: 2 \times 1}\\
				&=
				\begin{bmatrix}
					\big(\lambda_0 \cdot (1-X_1)+\lambda_1  \cdot X_1\big) \cdot T_1     \\
					\big(\lambda_0  \cdot (1-X_2)+\lambda_1 \cdot  X_2\big) \cdot T_2    \\
					\big((2-(X_1+X_2))  \lambda_0 +  (X_1+X_2)  \lambda_1\big)  T_3  \\
				\end{bmatrix} 
			\end{aligned}
		\end{equation}
		where $\Lambda_1=\lambda_0 \cdot (1-X_1)+\lambda_1  \cdot X_1$ and $\Lambda_2=\lambda_0 \cdot (1-X_2)+\lambda_1  \cdot X_2$ are two scalar functions of random variables $X_1$ and $X_2$, respectively.
		From the optimization point of view and in terms of \textit{sensor scheduling}, we are interested in finding the optimal time-allocation, $ (T_1,T_2,T_3) $, of total available time resource, $ T $, that would maximize the \textit{reward} i.e. either the mutual information or probability of total correct detections. We may say that we are interested in finding the diagonal matrix $ \mathcal{D} $, such that the mutual information $ I(X;Y) $ is maximized under time constraint $ T=T_1+T_2+T_3 $. Configuration matrix $ \mathcal{B} $ represents all possible sensing combinations and $ \Lambda $ is the Poisson rate parameter vector.  Vector $\mathcal{M}$ is the vector of intensities of the Poisson random variables $Y_1$, $Y_2$, and $Y_3$.
		
		Mathematically we may write 
		\begin{IEEEeqnarray*}{rCl}
			& \underset{T_1, T_2, T_3}{\text{max}} \: I(X_1, X_2 ; Y_1, Y_2, Y_3) \: \: \text{s.t.} \: \:  T_1+T_2+T_3=T. \label{e1} \IEEEyesnumber
		\end{IEEEeqnarray*} 
		We may rewrite the objective function 
		\begin{IEEEeqnarray*}{rCl}
			& \underset{T_1, T_2, T_3}{\text{max}} \: I(X_1, X_2 ; Y_1, Y_2, Y_3) \: \: \text{s.t.} \: \:  T_1+T+T_3=1 \label{e10} \IEEEyesnumber
		\end{IEEEeqnarray*} 
		where $0 \le T_1 \le 1$, $0 \le T_2 \le 1$ and $0 \le T_3 \le 1$. Without loss of generality we can take $T=1$ since the means of the observed $Y_i$ variables are the product of rate and time, and any change in the total available time can be reflected in the rates, or equivalently changing the time units.
		
		We extend our understanding by looking at the same problem from the \emph{detection theoretic} aspect. For this, we maximize the Bayesian probability of total correct detection, $P_d $, of hidden random vector $ X $ from observable random vector $ Y $, as 
		\begin{IEEEeqnarray*}{rCl}
			& \underset{T_1, T_2, T_3}{\text{max}} \: P_d \quad  \text{s.t.} \quad  T_1+T_2+T_3=1 \label{e9} \IEEEyesnumber
		\end{IEEEeqnarray*}
		and compare the results to formerly computed \emph{information theoretic} results. Simulations on empirical data are performed by varying different parameters involved in the model for further validation of computed results. 

		\section{Information Theoretic Description} \label{detofc}
		
		
		\subsection{Scalar Poisson channel} \label{form0}   
		\begin{figure}[t!] 
			\centering
			\includegraphics[width=.75\linewidth]{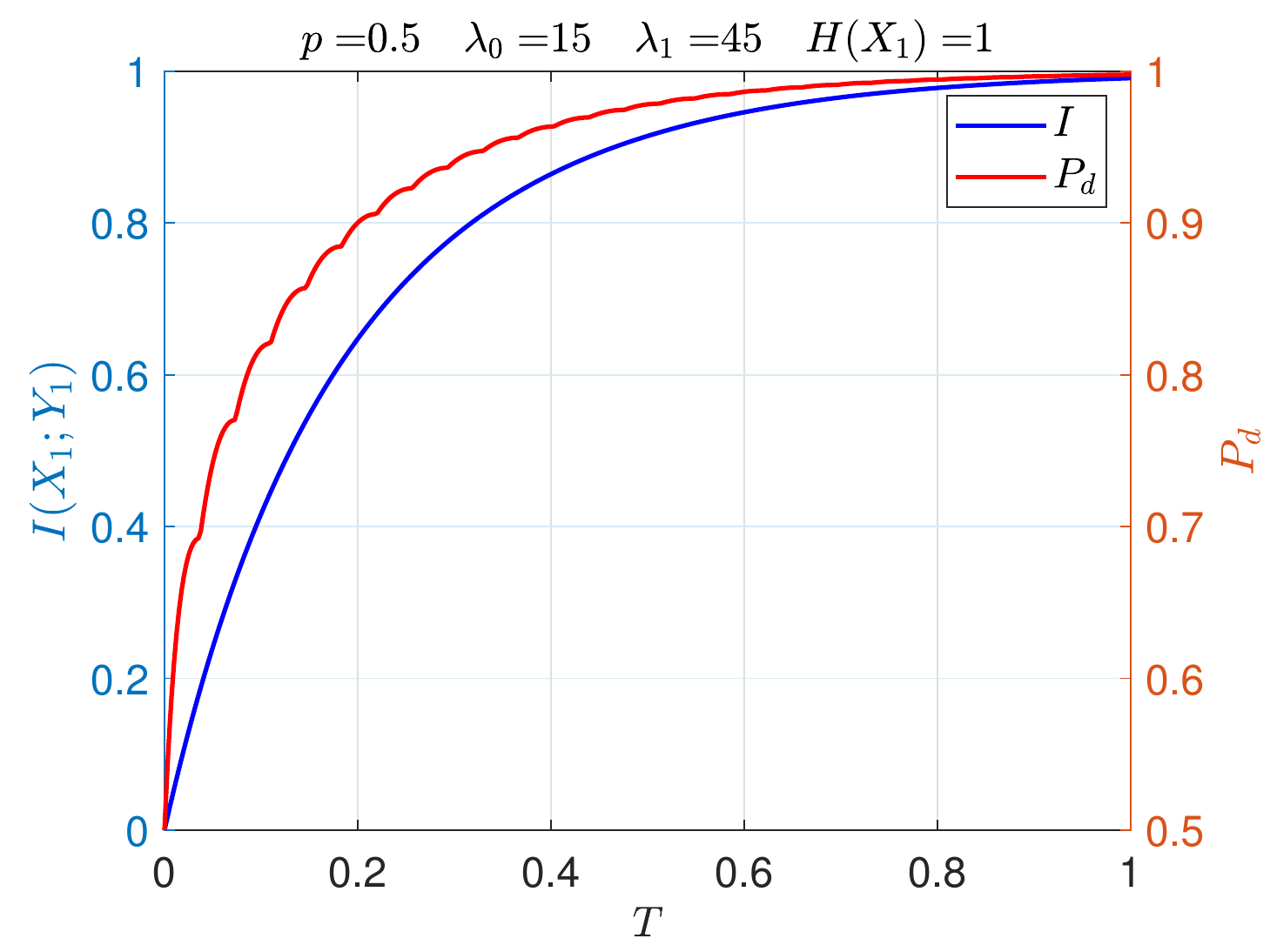}  
			\caption{Scalar Poisson channel: mutual information $ I(X_1;Y_1) $ and probability of total correct detections $P_d$ vs. time $T.  $ \todo[inline]{\texttt{\detokenize{poiss_mix.m}}                     }	}
			\label{f4}
		\end{figure}
		
		Firstly the scalar version of the Poisson channel is presented and then it is  extended to the vector version of our problem. We start with mutual information between a scalar Bernoulli random variable $ X_1 $ and $ Y_1 $ which is a scalar and two component Poisson mixture. The probability mass function of $ Y_1 $ is then given as
		\begin{IEEEeqnarray*}{rCl}
			f(y_1)	 &=& (1-p) \cdot  \operatorname{Poiss}(y_1;T \lambda_0) + p \cdot  \operatorname{Poiss}(y_1;T \lambda_1).
		\end{IEEEeqnarray*}
		The mutual information $ I $ can be written as
		\begin{IEEEeqnarray*}{rCl}
			I(X_1;Y_1) &=& H(Y_1)-H(Y_1|X_1)	
		\end{IEEEeqnarray*}
		where $ H(\cdot) $ is the Shannon entropy, which is defined as a discrete functional $ H: f \rightarrow - \sum\limits_{Y \in \mathcal{Y}}^{} f \cdot \operatorname{log}_2(f)   $  and $ f $ is the probability mass function of random variate $ Y $ with $ \mathcal{Y} $ as the corresponding support.
		We may write $ H(Y_1) $ as
		\begin{IEEEeqnarray*}{rCl}
			H(Y_1)  & = &  - \sum\limits_{y_1=0}^{\infty} \Big( [(1-p) \cdot  \operatorname{Poiss}(y_1;T \lambda_0) + p\\ && \cdot \> \operatorname{Poiss}(y_1;T \lambda_1)]  \cdot  \operatorname{log}_2[(1-p) \cdot\operatorname{Poiss}(y_1;T \lambda_0) \\&&+\> p\cdot  \operatorname{Poiss}(y_1;T \lambda_1)] \Big),
		\end{IEEEeqnarray*}
		and \begin{IEEEeqnarray*}{rCl}
			H(Y_1|X_1)	&=&  \sum\limits_{x_1 \in\mathcal X_1} \:  f(x_1) \: \sum\limits_{y_1=0}^{\infty} \: -f(y_1|x_1) \cdot \operatorname{log}_2[f(y_1|x_1)] \\
			&=& - \Big( (1-p) \sum\limits_{y_1=0}^{\infty} \: \operatorname{Poiss}(y_1;T \lambda_0) \\ && \> \cdot \operatorname{log}_2[\operatorname{Poiss}(y_1;T \lambda_0) ] + p \sum\limits_{y_1=0}^{\infty} \: \operatorname{Poiss}(y_1;T \lambda_1)  \\ && \>\cdot \operatorname{log}_2[\operatorname{Poiss}(y_1;T \lambda_1) ] \Big).
		\end{IEEEeqnarray*}
		In Fig. (\ref{f4}), for a scalar Poisson channel discussed above, $I(X_1;Y_1)$ and $P_d$ illustrates a monotonic relationship w.r.t $T$, as expected.
		\subsection{Vector Poisson channel}  \label{dermi}
		Mutual information between two random vectors can be defined as the difference between the total entropy in one random vector and the conditional entropy in the second random vector given the first vector. We write
		\begin{IEEEeqnarray*}{rCl}
			I(X;Y)&=& H(Y)-H(Y|X) \label{e5} \IEEEyesnumber
		\end{IEEEeqnarray*}  
		
		The conditional entropy $ H(Y|X) $ is calculated from the conditional probability mass functions $ f(y|X=[x_1 \quad x_2]^{\intercal}) $ defined as
		\begin{IEEEeqnarray*}{rCl}
			&&f(y|X=[0 \quad 0]^{\intercal})\nonumber\\ &&= \>\operatorname{Poiss}(y_1; \lambda_0 T_1) \cdot \operatorname{Poiss}(y_2; \lambda_0 T_2) \cdot  \operatorname{Poiss}(y_3; 2\lambda_0 T_3),  \\
			&&f(y|X=[0 \quad 1]^{\intercal})\nonumber\\ &&= \>\operatorname{Poiss}(y_1; \lambda_0 T_1) \cdot \operatorname{Poiss}(y_2; \lambda_1 T_2) \cdot  \operatorname{Poiss}(y_3; (\lambda_0 + \lambda_1) T_3),  \\
			&&f(y|X=[1 \quad 0]^{\intercal})\nonumber\\ &&= \>\operatorname{Poiss}(y_1; \lambda_1 T_1) \cdot \operatorname{Poiss}(y_2; \lambda_0 T_2) \cdot  \operatorname{Poiss}(y_3;(\lambda_1+\lambda_0) T_3),  \\
			&&f(y|X=[1 \quad 1]^{\intercal})\nonumber\\ &&= \>\operatorname{Poiss}(y_1; \lambda_1 T_1) \cdot \operatorname{Poiss}(y_2; \lambda_1 T_2) \cdot  \operatorname{Poiss}(y_3; 2\lambda_1 T_3). 
		\end{IEEEeqnarray*}
		The marginal probability mass function of $Y$ is then given as
		\begin{IEEEeqnarray*}{rCl}
			&&f(y)\nonumber\\ &&= (1-p)^2 \cdot f(y|X=[0 \quad 0]^{\intercal}) + p(1-p) \cdot \nonumber\\ && f(y|X=[0 \quad 1]^{\intercal})  +  p(1-p) \cdot f(y|X=[1 \quad 0]^{\intercal})  + \nonumber\\ && p^2 \cdot f(y|X=[1 \quad 1]^{\intercal}). \label{e05} \IEEEyesnumber
		\end{IEEEeqnarray*}
		Using the identity defined in (\ref{e5}) and the definition of entropy defined above, the mutual information $ I(X;Y) $ becomes
		\begin{IEEEeqnarray*}{rCl}
			&&I(X;Y)\nonumber\\ &&= \sum\limits_{y_1=0}^{\infty} \sum\limits_{y_2=0}^{\infty}  \sum\limits_{y_3=0}^{\infty} -f(y_1,y_2,y_3) \cdot \operatorname{log}_2[f(y_1,y_2,y_3)] )\nonumber\\ && \quad - \sum\limits_{X \in \mathcal X} f(x)  \Big[ \sum\limits_{y_1=0}^{\infty} \sum\limits_{y_2=0}^{\infty} \sum\limits_{y_3=0}^{\infty} -f(y_1,y_2,y_3|X) \nonumber\\&& \quad\cdot \operatorname{log}_2[f(y_1,y_2,y_3|X)]    \Big]. \label{e11} \IEEEyesnumber
		\end{IEEEeqnarray*}
		A complete expression for mutual information is given in Appendix \ref{mixy}. 
		
		Fig. (\ref{f0}) illustrates the concavity of mutual information as a function of $T_3$ (constrained problem), in a vector Poisson channel defined in (\ref{a}) and illustrated in Fig. (\ref{f1}), when three sensing times are varied according to $(T_1,T_2,T_3)=(\frac{1-T_3}{2},\frac{1-T_3}{2},T_3)$ and $0 \le T_3 \le 1$. It can also be seen that the two metrics $P_d$ and $I$ are maximizing at two different input argument values.
		\begin{theorem}
			$I(X_1,X_2;Y_1,Y_2,Y_3)$ is concave in $T_3=0$ plane.
		\end{theorem}
		\begin{IEEEproof}
			\\ $I(X_1,X_2;Y_1,Y_2,Y_3)\Big|_{T_3=0}=I(X_1,X_2;Y_1,Y_2)$ 
			\\By chain rule of mutual information:
			\begin{IEEEeqnarray*}{rCl}
				I(X_1,X_2;Y_1,Y_2)=&& \>I(X_1,X_2;Y_1)+I(X_1,X_2;Y_2|Y_1) \\  
				=&& I(X_1;Y_1) + I(X_2;Y_2). \label{e6} \IEEEyesnumber
			\end{IEEEeqnarray*}
			We note in (\ref{e6}) that each $ I(X_i;Y_i) $ is solely a function of $ T_i $ and also concave in it \cite[p. 1306]{atar2012mutual}\cite{phdfahad}. We further know that sum of concave functions is a concave function. Therefore $I(X_1,X_2;Y_1,Y_2)$ is concave in $T_1$ and $ T_2$ when $T_3=0$. This concludes the proof.
		\end{IEEEproof}
		\begin{theorem}
			$I(X_1,X_2;Y_1,Y_2,Y_3)$ is symmetric in variables $ T_1 $ and $ T_2 $.
		\end{theorem}
		\begin{IEEEproof}
			Mutual information $I(X_1,X_2;Y_1,Y_2,Y_3)$ given in appendix \ref{mixy} is invariant under any permutation of variables $ T_1 $ and $ T_2 $. That means interchanging the two variables leaves the expression unchanged.
		\end{IEEEproof}
		If we further expand the expression for mutual information between $ X $ and $ Y $ using chain rule \cite{yeung2008information} \cite{cover2012elements} as, \begin{IEEEeqnarray*}{rCl}
			&&I(X_1,X_2;Y_1,Y_2,Y_3)\nonumber\\ &&= \> I(X_1,X_2;Y_1) + I(X_1,X_2;Y_2|Y_1) + I(X_1,X_2;Y_3|Y_1,Y_2)   \\
			&&= \> I(X_1;Y_1)+\overbrace{I(X_2;Y_1|X_1)}^{0} + \overbrace{I(X_1;Y_2|Y_1)}^{0} +\\ &&\quad \> \overbrace{I(X_2;Y_2|Y_1,X_1)}^{I(X_2;Y_2)}  + I(X_1,X_2;Y_3|Y_1,Y_2)   \\
			&&= \> I(X_1;Y_1) +I(X_2;Y_2) + I(X_1,X_2;Y_3|Y_1,Y_2).\label{e3} \IEEEyesnumber
		\end{IEEEeqnarray*} 
		\begin{figure}[t!] 
			\centering
			\includegraphics[width=.75\linewidth]{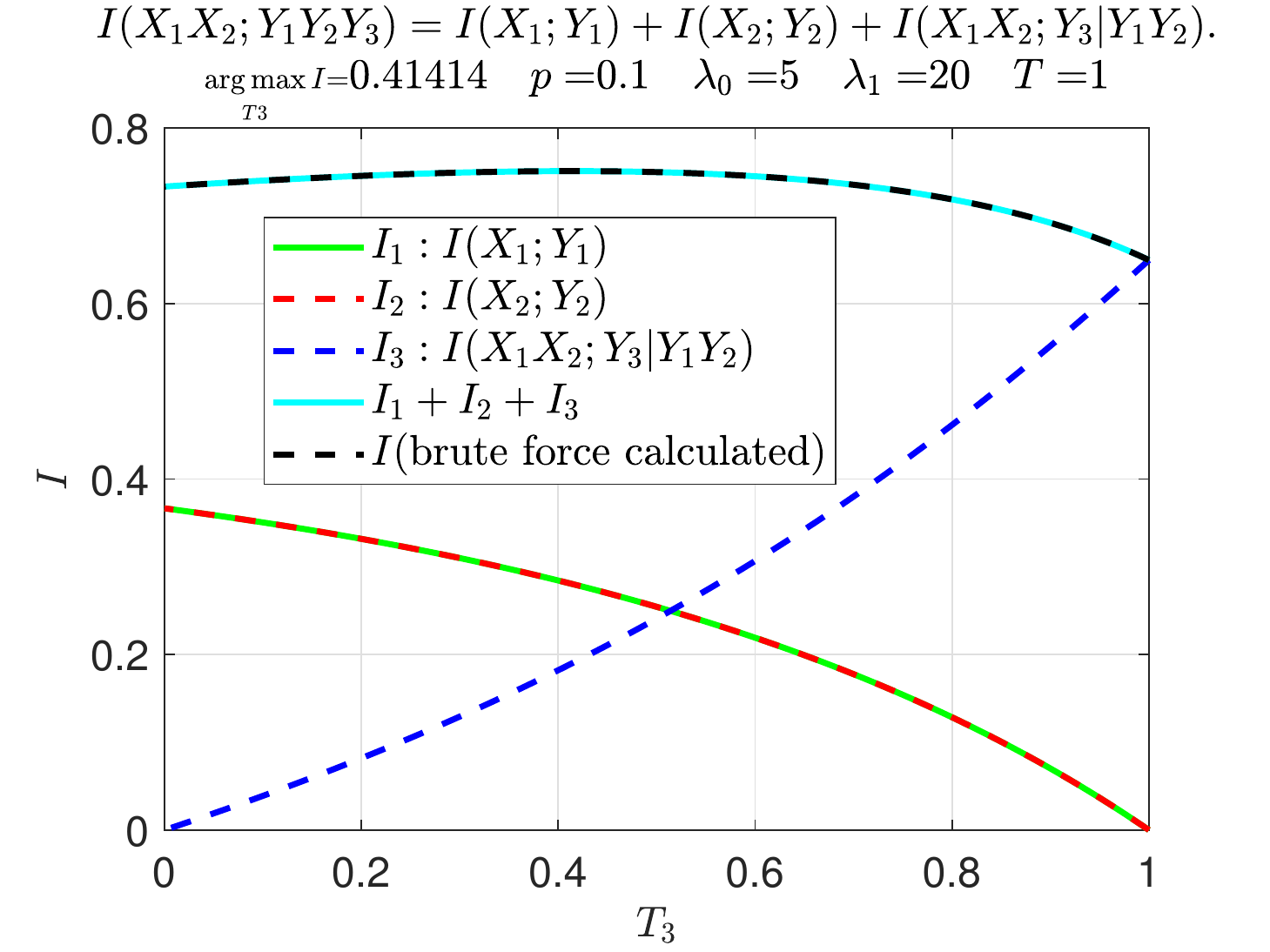}  
			\caption{Concavity and convexity of three terms in $I(X_1 X_2;Y_1 Y_2 Y_3)=I(X_1;Y_1)+I(X_2;Y_2)+I(X_1 X_2;Y_3|Y_1  Y_2)$ when  $ (T_1,T_2,T_3):=(\frac{1-T_3}{2},\frac{1-T_3}{2},T_3)$  such that $ 0 \le T_3 \le 1$.
				\todo[disable,inline]{Con\_3rdTerm.m     \newline     Con\_MI\_2.m                 }	}
		\label{f20}
	\end{figure}   
	Note that in (\ref{e3}) that the first and second terms are indeed concave in $ (T_1,T_2,T_3) $. The third term, when computed, exhibits non-concavity in the respective domain. Fig. (\ref{f20}) illustrates this when mutual information is plotted along the line $ (T_1,T_2,T_3):=(\frac{T-T_3}{2},\frac{T-T_3}{2},T_3)$  parameterized by $ 0 \le T_3 \le T $. It is interesting to note that the sum of three terms exhibits concavity, irrespective of the fact that the third term is non-concave. However, analytical investigation of the functional properties of this third term, $ I(X_1 X_2;Y_3|Y_1  Y_2) $, is not done in this work. 
	
	We computed mutual information between hidden random vector $ X $ and observable random vector $ Y $ for diverse set of intensities $ \lambda_0 $ and $ \lambda_1 $ along with different priors and total available time, $ T=1 $, to observe the concavity of $ I(X_1,X_2;Y_1,Y_2,Y_3) $ in $ T_1 $, $ T_2 $ and $ T_3 $. While we do not have a proof that the objective function is concave in a linear time constraint for the $I$ metric, we have carried out extensive computational experiments and we have yet to observe a single case in which non-concavity is observed. Based on our observation we consistently noted the concave property of mutual information which led us to propose the following conjecture.
	
		\begin{conjecture}
			If $ X \equiv [ X_1 \: X_2 ]^{\intercal} $ be a non-negative random vector where $ X_1 $ and $ X_2 $ are mutually independent and identical distributed. Let random vector $ Y \equiv [ {Y_1} \:  {Y_2} \:  {Y_3}]^{\intercal} \in \mathbb{Z}_+^3 $  ,  jointly distributed with $ X $ such that conditional law is given as in (\ref{a}) then $ I(X_1,X_2;Y_1,Y_2,Y_3) $ is concave in $ (T_1,T_2,T_3) $ under time constraint  $ T=T_1+T_2+T_3 $.
		\end{conjecture}
		
		
		\begin{corollary}[Feasible region for optimal solution]
			The two properties of mutual information  \emph{concavity} (if true) and \emph{symmetry} (proved), guarantee that \emph{maxima} must occur at the line of symmetry $ (T_1,T_2,T_3) := (\frac{T-\alpha}{2},\frac{T-\alpha}{2},\alpha)$ where $0 \le \alpha \le T$.
		\end{corollary}
		
		One of the immediate consequences of exploiting \emph{symmetry} and \emph{concavity} of the problem is that it would reduce the search space for the optimal solution from two dimensions to one dimension.
		
		\textit{Claim} : Mutual information $I(X_1, X_2;Y_1, Y_2, Y_3)$ in general is not symmetric in $ p $ around $p=0.5$ for fixed time proportions $(T_1,T_2,T_3)$ for a conditionally multivariate Poisson $(Y_1,Y_2,Y_3)$ given $(X_1,X_2)$.
		
		\begin{IEEEproof}[Example]
			For a scalar random variable $X_1$ and conditional Poisson variable $Y_1$ with scaling factor $T$, we have equation $(115)$ on page $(10)$ of  \cite{guo2008mutual}, given below
			\begin{IEEEeqnarray*}{rCl}
				\frac{d}{d T} I(X_1;Y_1) \Big |_{T=0}=E[\Lambda_1 \cdot \operatorname{log}{\Lambda_1}]-E[\Lambda_1]\cdot \operatorname{log}(E[\Lambda_1]). \label{e00} \\ \IEEEyesnumber 
			\end{IEEEeqnarray*}	
			where $\Lambda_1=\lambda_0 \cdot (1-X_1)+\lambda_1  \cdot X_1$. When we consider a random vector $X$ (instead of scalar random variable as above), and take distribution on $X$ as given in (\ref{w8}).  
			Dividing variable total time $T$ equally in counting arrivals from $X_1$ and $X_2$ separately: $T_1=T_2=\frac{T}{2}$ and $T_3=0$, we have
			\begin{IEEEeqnarray*}{rCl}
				I(X_1,X_2;Y_1,Y_2,Y_3) &\Big |&_{\big(T_1=\frac{T}{2},T_2=\frac{T}{2},T_3=0 \big)}= \\ \> && 2 \cdot I(X_1;Y_1) \Big|_{\big(T_1=\frac{T}{2}\big)} \label{L2}  \IEEEyesnumber 
			\end{IEEEeqnarray*}
			
			Let $T$ be a free variable in (\ref{L2}), we have
			\begin{IEEEeqnarray*}{rCl}
				&&\frac{d}{dT} \Bigg(I(X_1,X_2;Y_1,Y_2,Y_3) \Big|_{\big(T_1=\frac{T}{2},T_2=\frac{T}{2},T_3=0 \big)}  \Bigg) \Bigg|_{T=0}= \IEEEnonumber\\ &&(1-p) \cdot\lambda_0 \cdot \operatorname{log}_2(\lambda_0)+p  \cdot  \lambda_1 \cdot \operatorname{log}_2(\lambda_1)- \\ && \Big((1-p) \cdot  \lambda_0+  p \cdot\lambda_1 \Big) \cdot \operatorname{log}_2\Big((1-p) \cdot \lambda_0+p \cdot \lambda_1 \Big) \label{L3} \IEEEyesnumber
			\end{IEEEeqnarray*}
			Equation (\ref{L3}) is concave in $p$
			\begin{IEEEeqnarray*}{rCl}
				&& \frac{d^2}{dp^2} \Bigg(\frac{d}{dT} \bigg(I(X_1,X_2;Y_1,Y_2,Y_3) \Big|_{\big(T_1=\frac{T}{2},T_2=\frac{T}{2},T_3=0 \big)}  \bigg) \Bigg|_{T=0} \Bigg) \\ \> &&=  -\frac{(\lambda_0- \lambda_1 )^2}{\operatorname{Ln}(2) \cdot \big((1-p) \lambda_0 +p \cdot \lambda_1\big)} <0 \label{L4} \IEEEyesnumber
			\end{IEEEeqnarray*}
			Maxima of equation (\ref{L3}) occurs at $p=\frac{4-e}{e}$ when $\lambda_0=2$ and $\lambda_1=4.$ Since (\ref{L3}) is concave in $p$, then if it was symmetric, maxima must had occurred at $p=0.5$. 
		\end{IEEEproof}
		\section{Detection Theoretic Description} \label{detd}
		\subsection{Maximization of probability of total correct detections in sensing time intervals} \label{adoptcri}
		In previous section, we discussed and presented the metric of mutual information $ I $ between hidden random vector $ X, $ and observable vector $ Y $. Here we approach the original sensing problem as a multi-hypothesis detection problem and find the optimal solution by minimizing the Bayesian risk \cite[pp.220]{schonhoff2006detection} in sensing times i-e $ (T_1,T_2,T_3) $. We define Bayes risk $ r $ as
		\begin{IEEEeqnarray*}{rCl}
			r &=& (1-p)^2 \Big [P_{00 \:| \: 00} \: C_{00 \:| \: 00} + P_{01 \:| \: 00} \: C_{01 \:| \: 00}  \nonumber\\ && \>+P_{10 \:| \: 00} \: C_{10 \:| \: 00}+ P_{11 \:| \: 00} \: C_{11 \:| \: 00} \Big] + p(1-p)\nonumber\\ && \>\Big [P_{00 \:| \: 01} \: C_{00 \:| \: 01} + P_{01 \:| \: 01} \: C_{01 \:| \: 01} + P_{10 \:| \: 01} \: C_{10 \:| \: 01}\nonumber\\ && \>+ P_{11 \:| \: 01} \: C_{11 \:| \: 01} \Big]+  p(1-p) \Big [P_{00 \:| \: 10} \: C_{00 \:| \: 10} \nonumber\\ && \>+ P_{01 \:| \: 10} \: C_{01 \:| \: 10} + P_{10 \:| \: 10} \: C_{10 \:| \: 10}+ P_{11 \:| \: 10} \: C_{11 \:| \: 10} \Big] \nonumber\\ && \> +p^2 \Big [P_{00 \:| \: 11} \: C_{00 \:| \: 11} + P_{01 \:| \: 11} \: C_{01 \:| \: 11} + P_{10 \:| \: 11} \: C_{10 \:| \: 11}\nonumber\\ && \>+ P_{11 \:| \: 11} \: C_{11 \:| \: 11} \Big],
		\end{IEEEeqnarray*}
		where $ P_{kl \:| \: ij}  $ is the probability that $ X=[x^i_1 \: x^j_2]^{\intercal} $ is true while decision $ X=[x^k_1 \: x^l_2]^{\intercal} $ is made; similarly for $ C_{kl \:| \: ij} $. Setting all costs for which $ [x^i_1 \: x^j_2]^{\intercal} \neq [x^k_1 \: x^l_2]^{\intercal} $ to one and  $[x^i_1 \: x^j_2]^{\intercal} = [x^k_1 \: x^l_2]^{\intercal}  $ to zero, we have 
		\begin{IEEEeqnarray}{rCl}
			r &=& (1-p)^2 \Big [P_{01 \:| \: 00} \: + P_{10 \:| \: 00} \: + P_{11 \:| \: 00} \: \Big] +  p(1-p) \nonumber\\ && \>\Big [P_{00 \:| \: 01} \:  +  P_{10 \:| \: 01} \:+ P_{11 \:| \: 01} \:  \Big]+ p(1-p)\nonumber\\ && \> \Big [P_{00 \:| \: 10} \:  + P_{01 \:| \: 10} \: + P_{11 \:| \: 10} \:  \Big]+   p^2\nonumber\\ && \> \Big [P_{00 \:| \: 11} \:  + P_{01 \:| \: 11} \:  + P_{10 \:| \: 11} \:  \Big]. \label{e2}
		\end{IEEEeqnarray}  
		We are interested in minimizing this Bayes risk $ r $ in $ (T_1,T_2,T_3)$ i-e
		\begin{IEEEeqnarray*}{rCl}
			& \underset{T_1, T_2, T_3}{\text{min}} \: r \quad \text{s.t.} \: \:  T_1+T_2+T_3=1. \label{e4} \IEEEyesnumber
		\end{IEEEeqnarray*} 
		Note that while minimizing $ r $ in $ (T_1,T_2,T_3)$, the decisions boundaries would be changing accordingly and become function of $ (T_1,T_2,T_3)$.
		Equivalently, we may say that
		\begin{IEEEeqnarray*}{rCl}
			& \underset{T_1, T_2, T_3}{\text{max}} \: P_d \quad  \text{s.t.} \: \:  T_1+T_2+T_3=1 \label{e8} \IEEEyesnumber
		\end{IEEEeqnarray*}
		where $ P_d $ is probability of total correct detections, $ P_d=1-r. $ In the next section we compute $I$ ans $P_d$, for different channel parameters, given in Appendices \ref{mixy} and \ref{pd}, respectively.
		\section{Computational and Simulation Results} \label{comand}
		
		The primary purpose of performing the computations and simulations is twofold: the first is to see under what circumstances which of the sensing method: individual, joint or hybrid sensing is optimal since analytical solutions for the two objective functions do not exist; and the second is to investigate whether the two metrics $I$ and $P_d$ leads to the same or different maximizing input arguments. The objective function, defined in (\ref{e1}) is observed to be a convex optimization problem when a pre-defined channel scaling matrix structure is imposed. For all computational purposes, we always approximate the Poisson mass function by truncating it; truncation is done by a rectangular window that extends from lower limit $ 0 $ to the upper limit where Poisson pmf drops in value below double precision machine epsilon ($ \epsilon=2^{-53} $) as given in Appendix \ref{mi}.
		\begin{figure}[t!] 
			\centering
			\includegraphics[width=.75\linewidth]{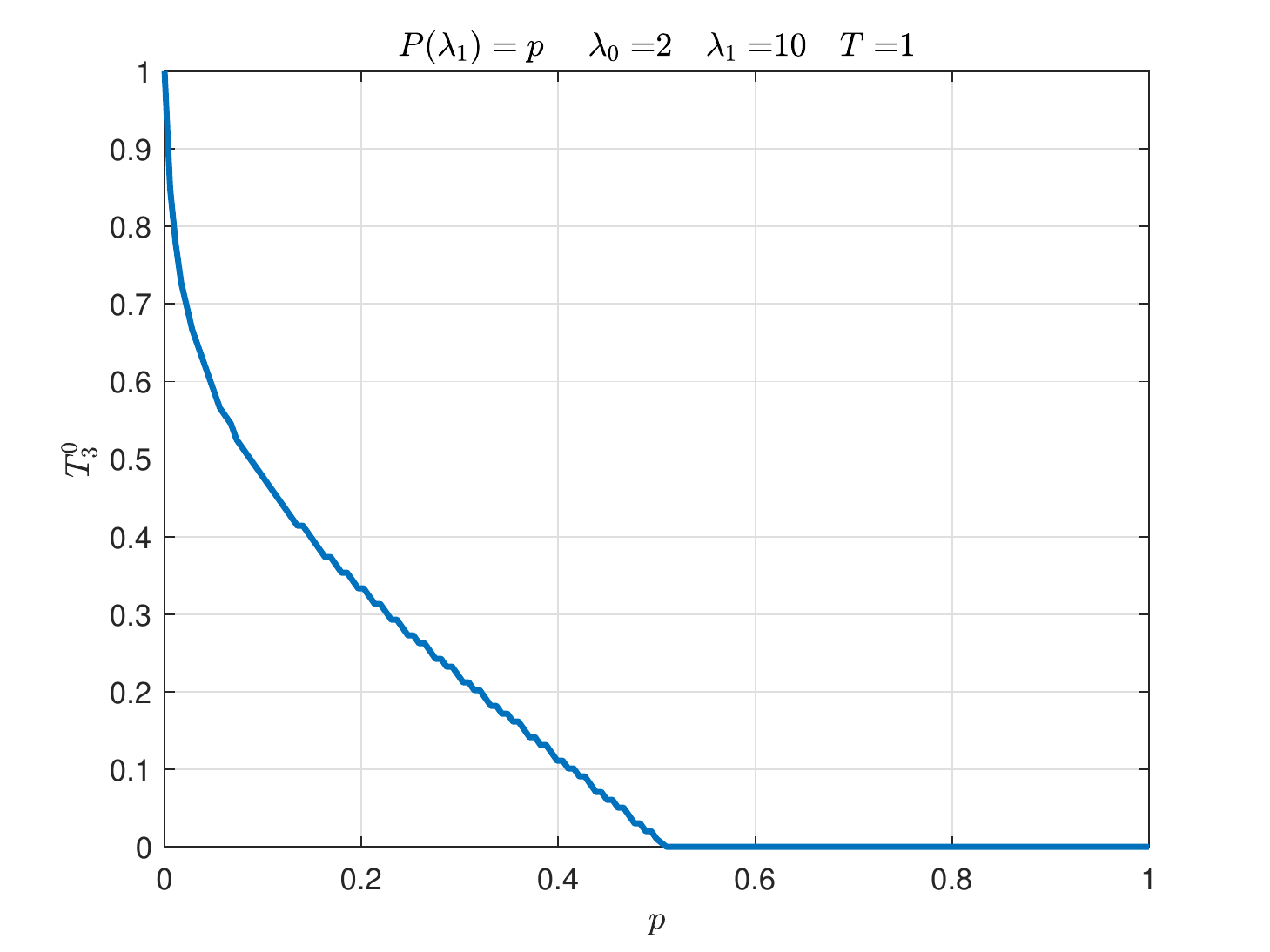}  
			\caption{Optimal time $ T_3^0 $ vs. prior probability $ p $ with $T_1=T_2 $ and $ T_1+T_2+T_3=T $, considering $I$ as a metric.
				\todo[disable,inline]{T3vsMI\_paper1     \newline     MI\_2.m                 }	}
		\label{f8}
	\end{figure}
	
	Fig. (\ref{f8}) illustrates how the optimal input argument $T_3$ moves from $0$ to $1$ as prior $0 <p <1$ is varied while input intensities are held fixed. Ninety linearly spaced discrete points in the range $(0,0.5]$ are first taken for prior $p$ and then another ten linearly spaced points are taken in the interval $(0.5,1)$, constituting a total of hundred points of $p$ for computations. For every value of  prior $p$; another hundred linearly spaced points for time proportion $T_3$ are taken in the closed interval $[0,1]$ such that $T_1=T_2=\frac{1-T_3}{2}$ and $T_1+T_2+T_3=1$. At every instant of $p$, hundred values of $I$ are computed corresponding to hundred time proportion. Maximum value of $I$ is then sought and the corresponding value of $T_3$ is plotted against prior $p$. The plot signifies that if the prior $p$ is closer to $0$ then all time $T$ be invested in joint sensing whereas as prior increases from $0$ towards $1$, individual sensing progressively starts to take more and more  proportion of total available time $T$. This phenomena is further explored in scatter plots where input intensities are varied too.
	
\begin{figure*}[ht]
		\begin{subfigure}{.49\textwidth}
			\includegraphics[width=\linewidth]{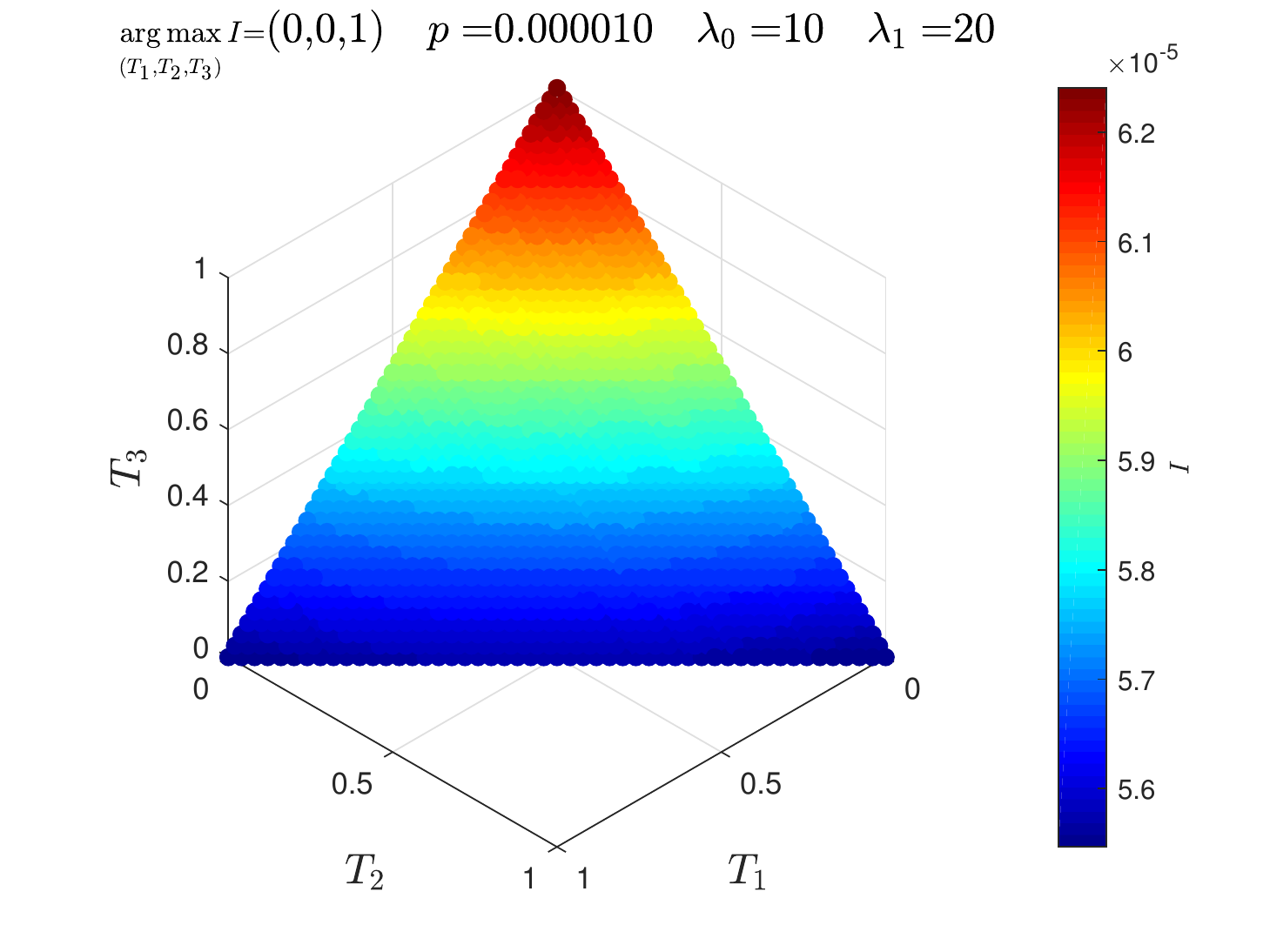}
			\caption{ }
			\label{fig2a}
		\end{subfigure} 
		\begin{subfigure}{.49\textwidth}
			\includegraphics[width=\linewidth]{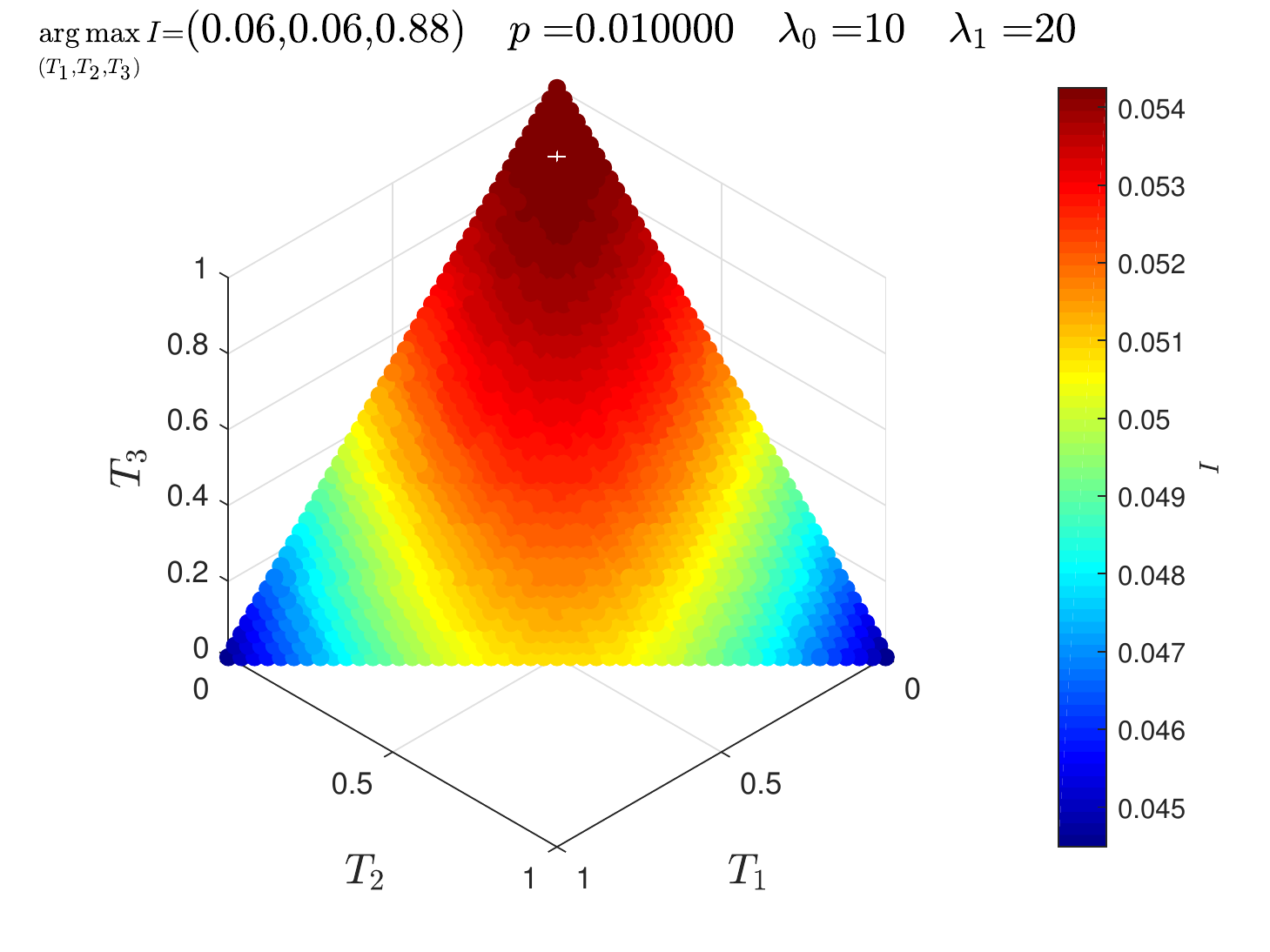}
			\caption{ }
			\label{fig2b}
		\end{subfigure} \\%
		\begin{subfigure}{.49\textwidth}
			\includegraphics[width=\linewidth]{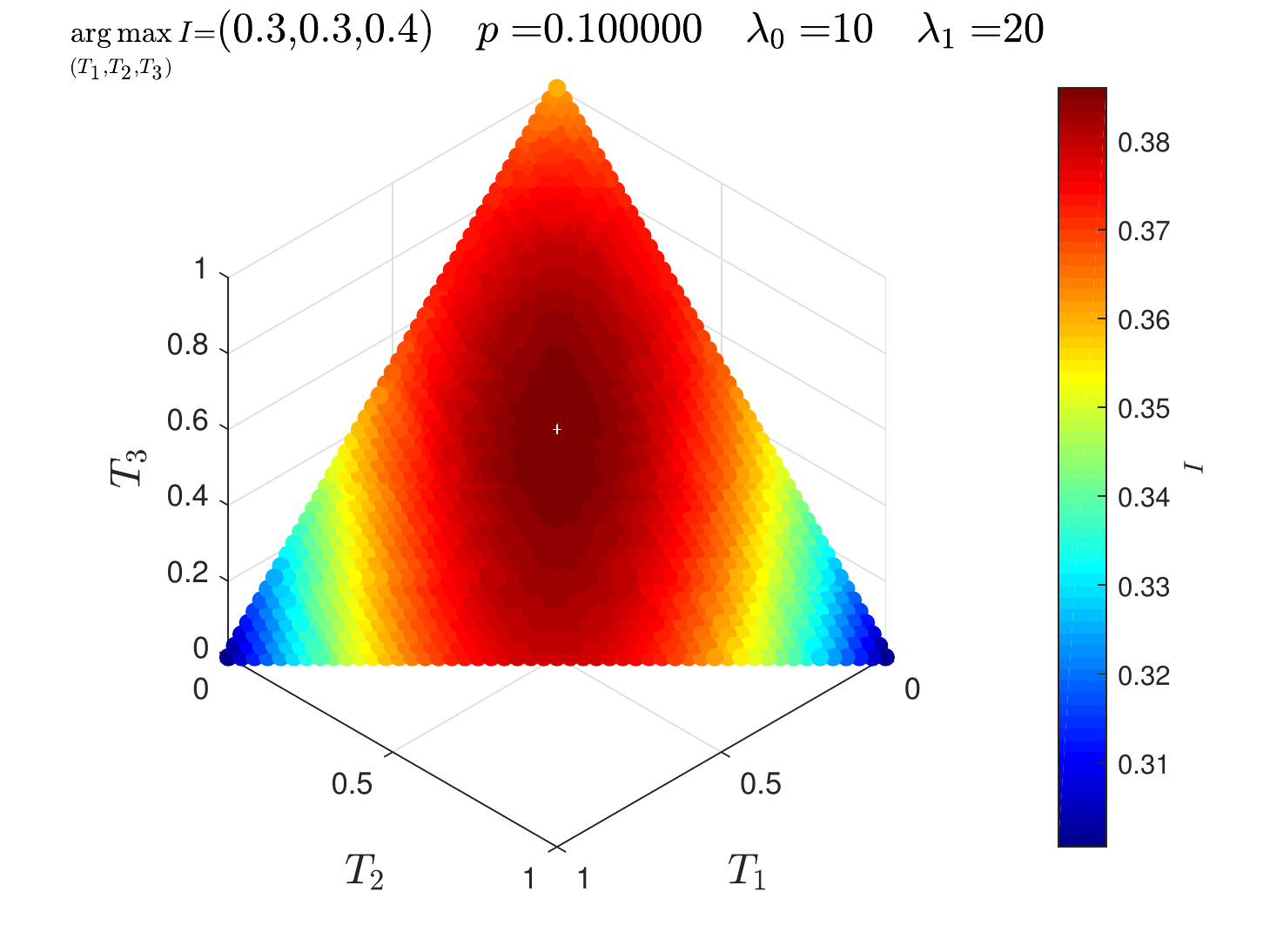}
			\caption{ }
			\label{fig2c}
		\end{subfigure} %
		\begin{subfigure}{.49\textwidth}
			\includegraphics[width=\linewidth]{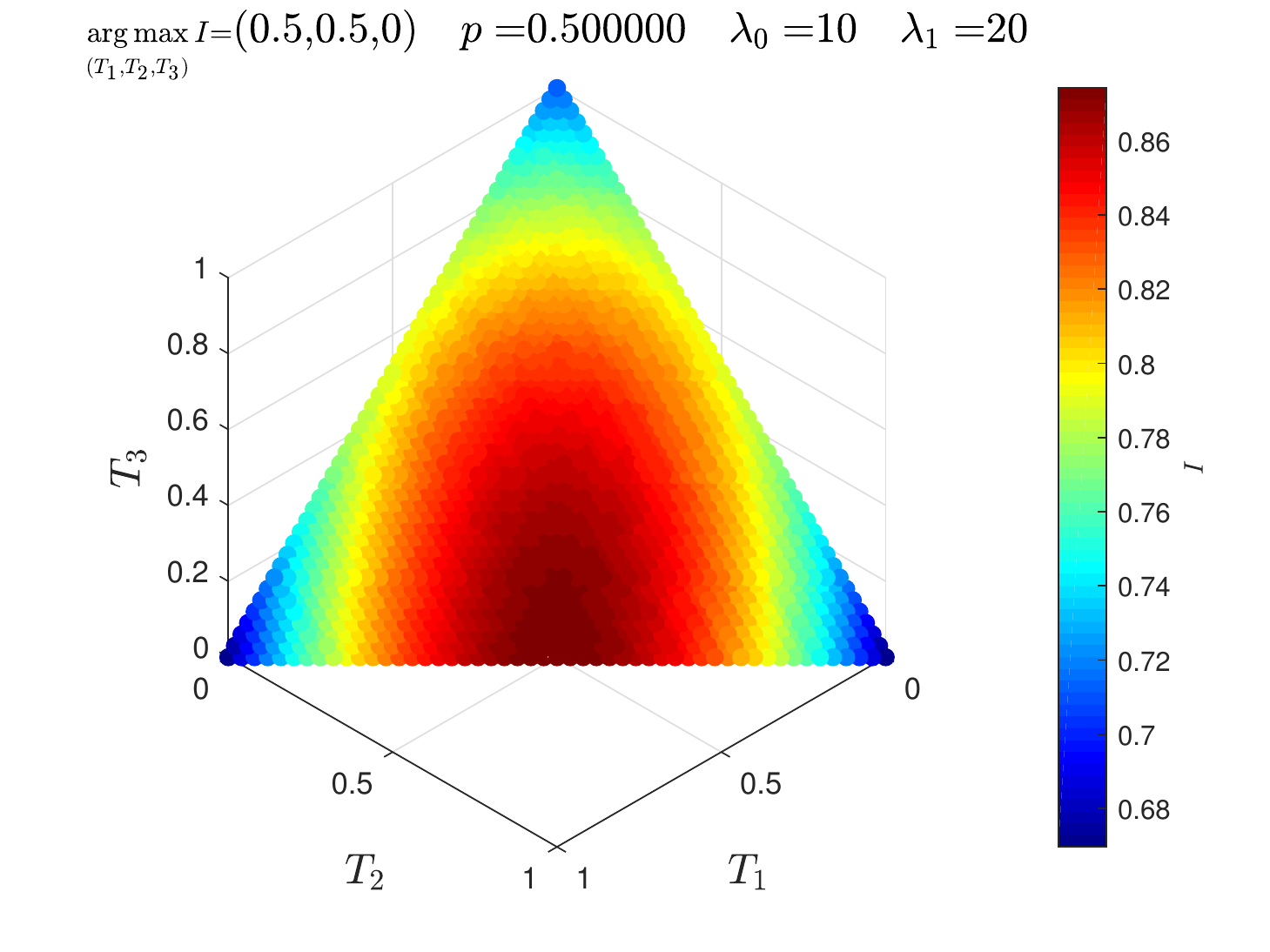}
			\caption{ }
			\label{fig2d}
		\end{subfigure}
		\begin{subfigure}{.49\textwidth}
			\includegraphics[width=\linewidth]{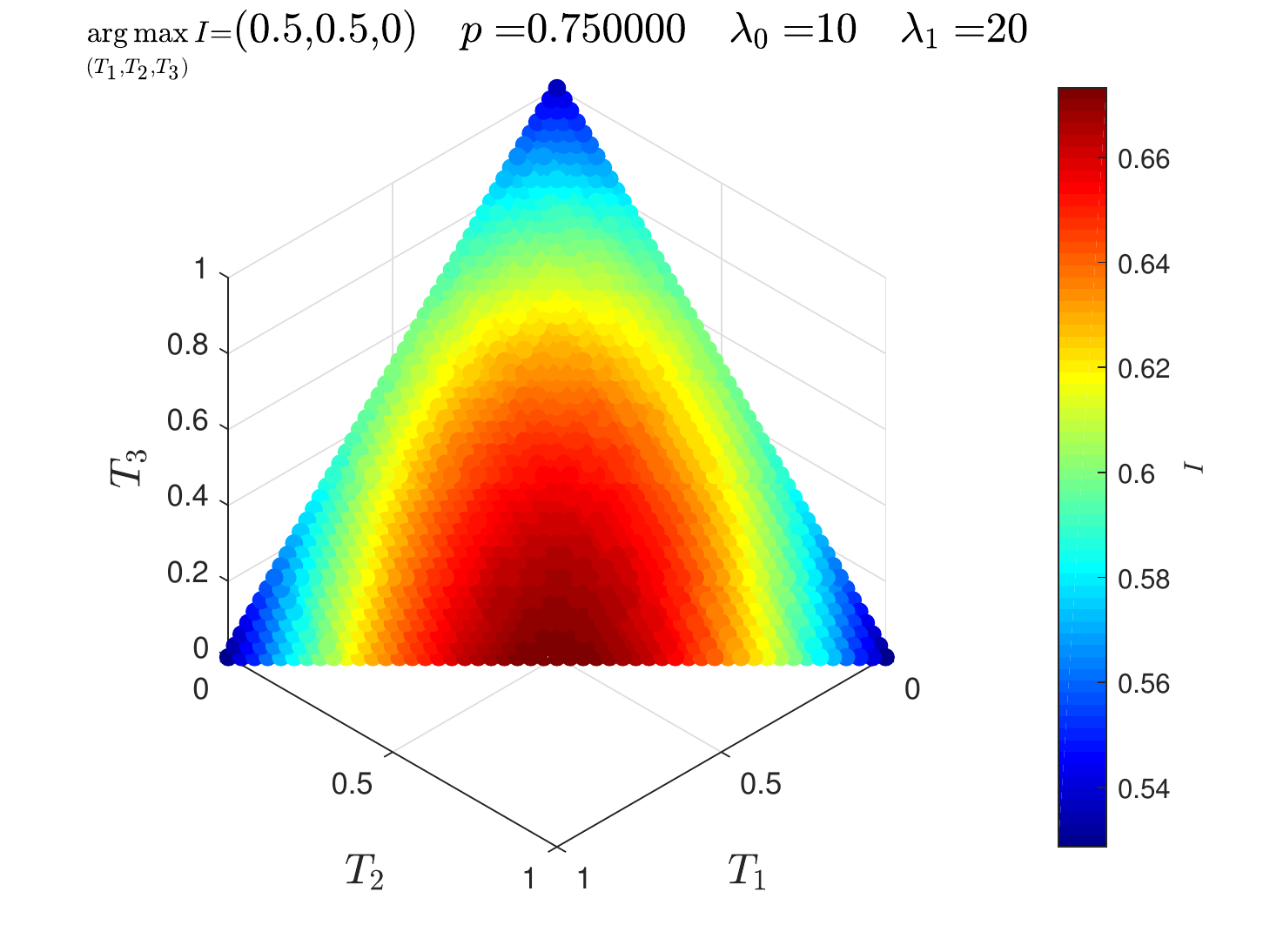}
			\caption{ }
			\label{fig2e}
		\end{subfigure} %
		\begin{subfigure}{.49\textwidth}
			\includegraphics[width=\linewidth]{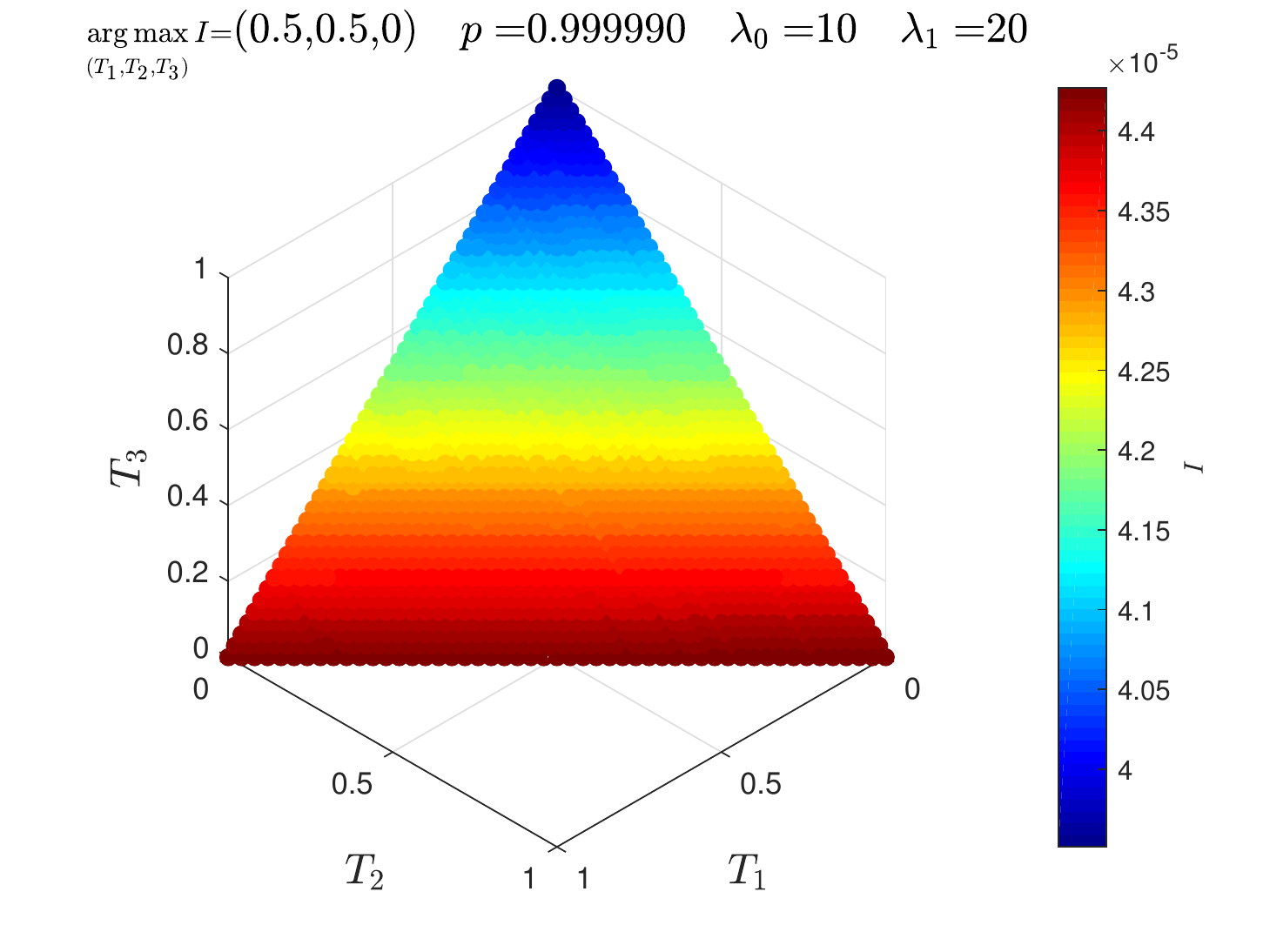}
			\caption{ }
			\label{fig2f}
		\end{subfigure} %
		\caption{$ I(X;Y) $ vs. $ (T_1,T_2,T_3) $ under time constraint $ T_1+T_2+T_3=1 $ for $ \lambda_0=10 $, $ \lambda_1=20 $, $ T=1 $ and varying \emph{prior} probability $ p $. It can be seen from (a)-(f) that as prior $ p $ varies from $ 0.00001  $ to $ 0.5$, the optimal solution drifts from $(0,0,1) $ to $(0.5,0.5,0) $ and stays there as $p$ is varied further from $0.5$ to $0.99999$ along the line of symmetry $ (T_1,T_2,T_3) := (\frac{1-\alpha}{2},\frac{1-\alpha}{2},\alpha)$ where $0 \le \alpha \le 1 .$ \todo[disable,inline]{\texttt{\detokenize{Two_bin_paper1.m}}
				\newline   	\texttt{\detokenize{MI_2.m}}                     }}
		\label{f2}
	\end{figure*}

	In Fig. (\ref{f2}), ternary plots of $I$ are shown as prior $ p $ is varied. The vertices of the equilateral triangle are at $(T_1,0,0)$,  $(0,T_2,0)$ and  $(0,0,T_3)$ with $T_1+T_2+T_3=1$. It can be seen that the distribution of time resource $T=1$ changes as $ p $ is varied. As $ p $ tends to $0$, the optimal value of $ T_3 $ tends to consume all the available time resource $ T $ leaving optimal values of $ T_1 $ and $ T_2 $ approaching to $ 0 $.  Mirror symmetry in arguments $ T_1 $ and $ T_2 $ can be observed. It can be seen further that the optimal argument always lies on the line that bisects the equilateral triangle into two right angle triangles i.e. $(T_1,T_2,T_3):=(\frac{1-T_3}{2},\frac{1-T_3}{2},T_3)$ where $0 \le T_3 \le 1$. 
	
\begin{figure}[h]
		\centering
		\begin{subfigure}[b]{.49\textwidth}
			\centering
			\includegraphics[width=1\linewidth,height=50mm]{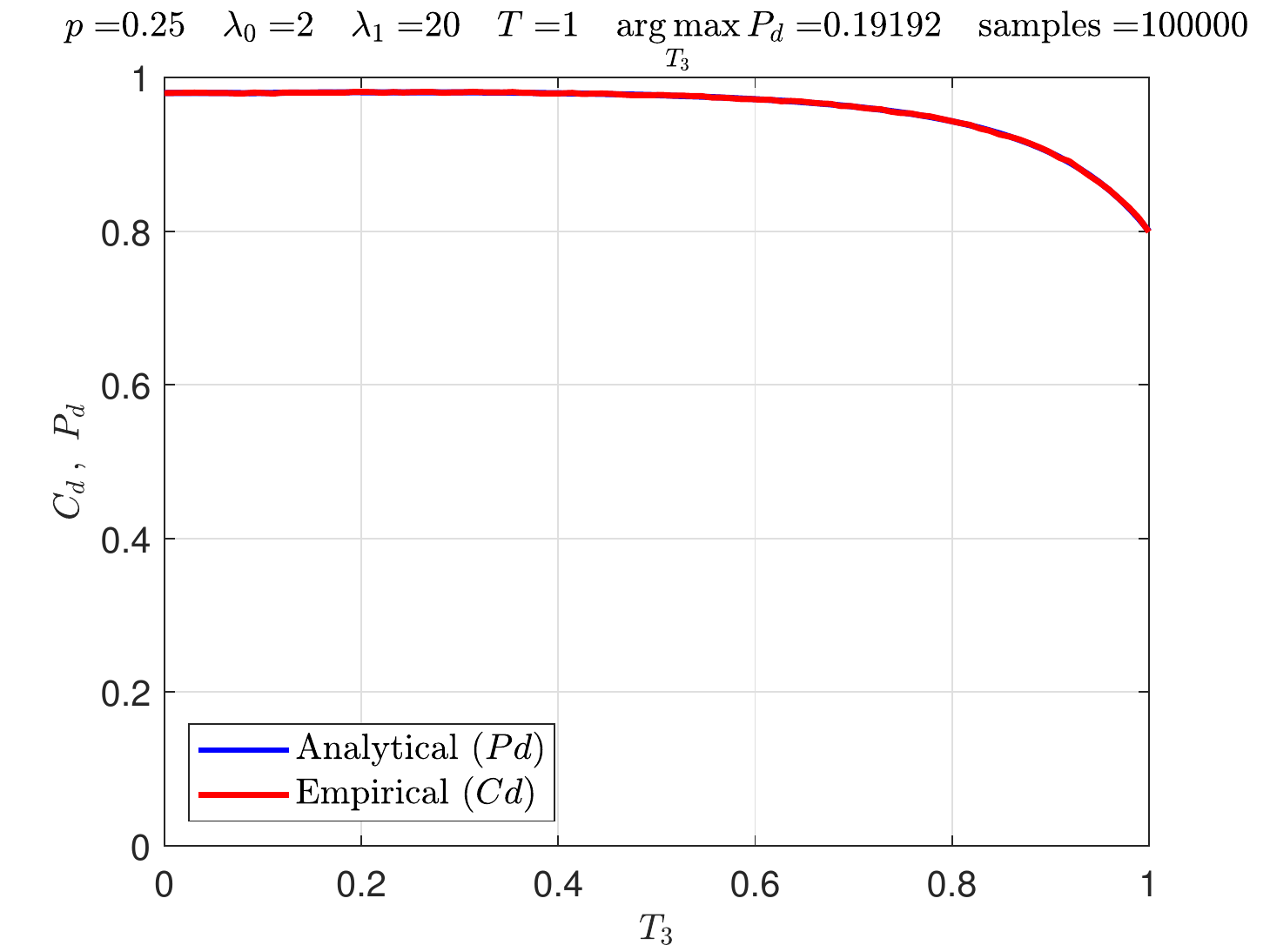}
			\caption{}
			\label{fig5a}
		\end{subfigure}
		\begin{subfigure}[b]{.49\textwidth}
			\centering
			\includegraphics[width=1\linewidth,height=50mm]{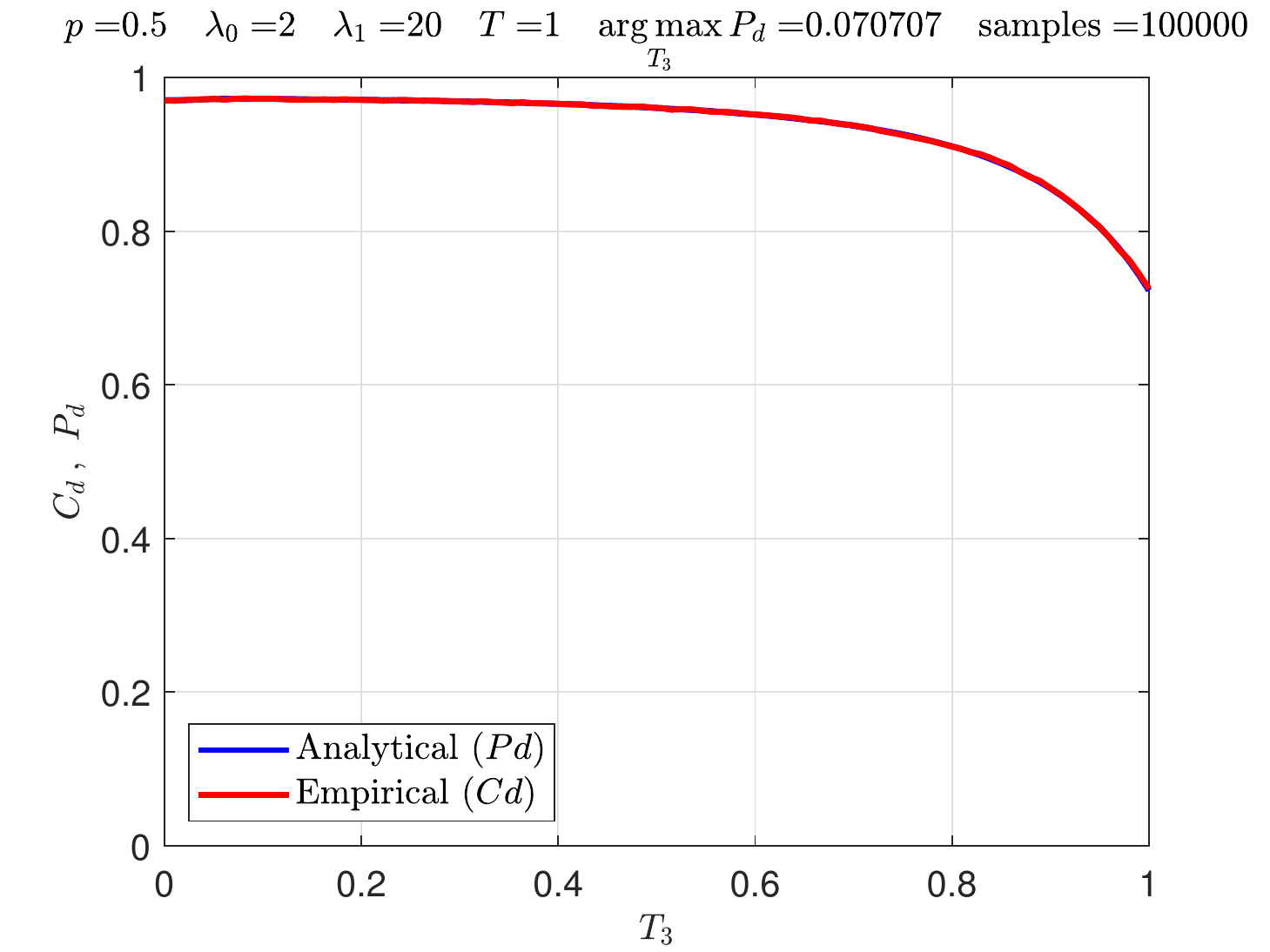}
			\caption{}
			\label{fig5b}
		\end{subfigure} %
		\begin{subfigure}[b]{.49\textwidth}
			\centering
			\includegraphics[width=1\linewidth,height=50mm]{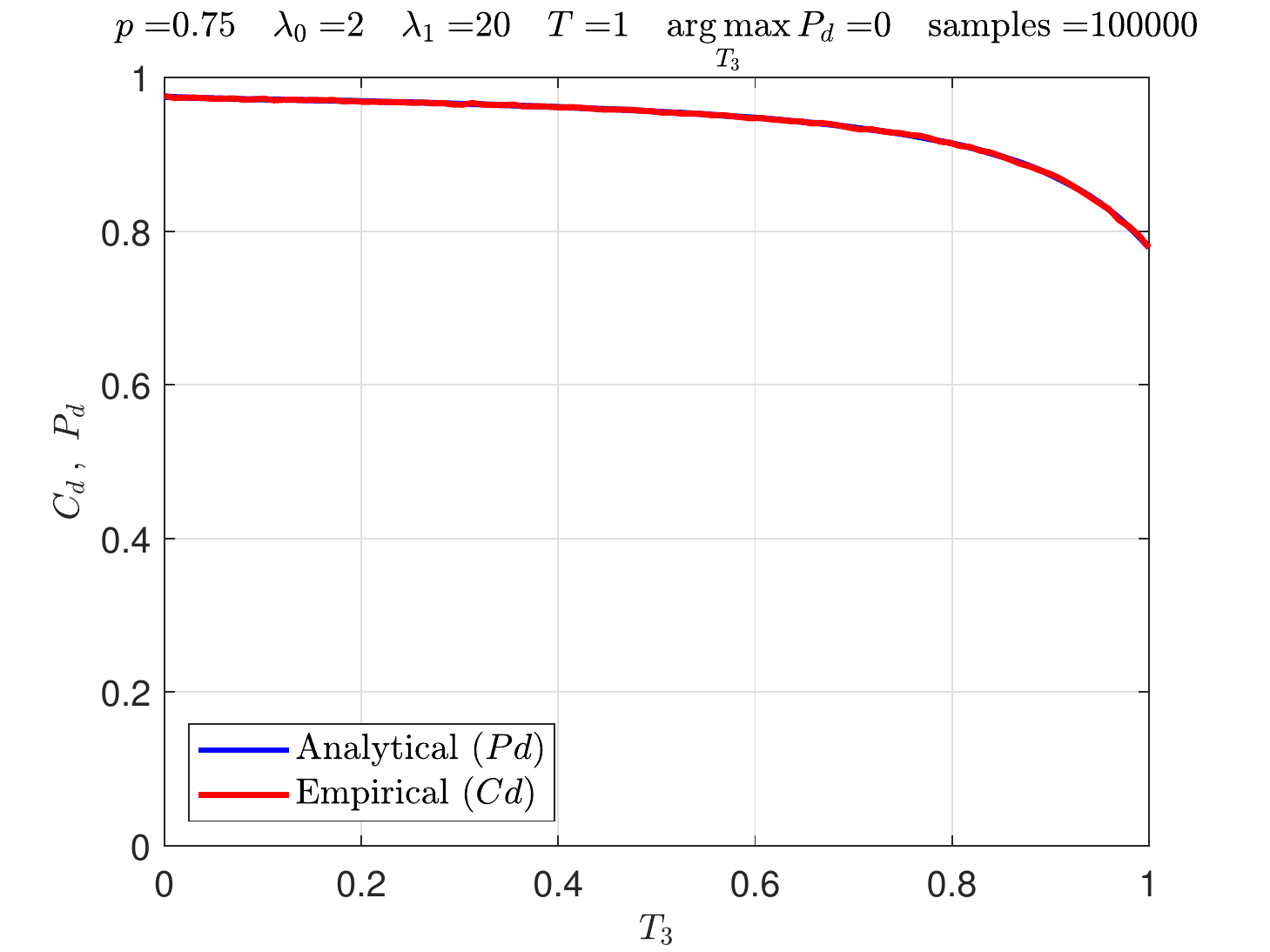}
			\caption{}
			\label{fig5c}
		\end{subfigure} %
		\begin{subfigure}[b]{.49\textwidth}
			\centering
			\includegraphics[width=1\linewidth,height=50mm]{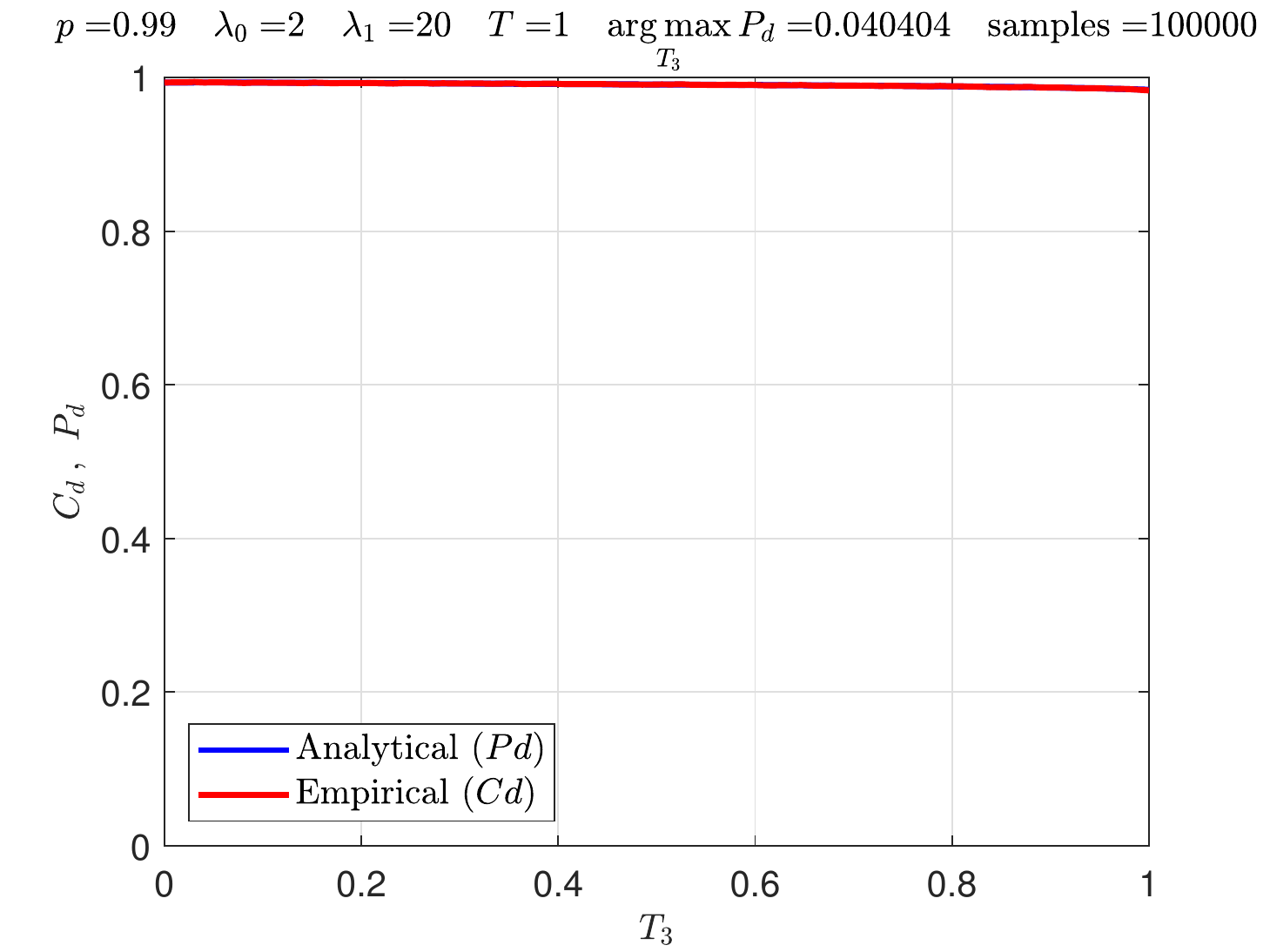}
			\caption{}
			\label{fig5d}
		\end{subfigure}
		\caption{Empirical Correct-decision rate $ C_d $ and analytical probability of total correct detections $ P_d $ vs. $ T_3 $ for \emph{prior} probabilities of $ 0.25,0.5,0.75 $ and $ 0.99. $  \todo[disable,inline]{\texttt{\detokenize{Pd_Cd_MI_FullSupport_paper1.m}}
				\newline   	\texttt{\detokenize{MI_2.m}}                     }}
		\label{f5}	
	\end{figure}

	A relationship between mutual information  and MAP (maximum a posteriori) detection for our problem is investigated in Fig. (\ref{f3}), in which Bayesian probability of total correct detections, $ P_d $, given in Appendix \ref{pd} is plotted against respective mutual information $ I $ given in Appendix \ref{mixy}. The $I$ and $P_d$ curves are not optimizing on the same input arguments. We further counter-checked our MAP detector performance with empirical means as shown in Fig. (\ref{f5}) by generating $10^5 $ samples from a multivariate Poisson mixture random vector $Y$, at each time instant $(T_1,T_2,T_3)$ for fixed given parameters: $p$, $\lambda_0$, $\lambda_1$. Total of $200$ linearly spaced time instants of $T_3$ are taken in the range $[0,1]$ and assuming $T_1=T_2=\frac{1-T_3}{2}$. The samples are taken in the same proportion (of $10^5$) from each of the four components of mixture $Y$  as their respective prior probabilities suggest. Each of the samples is then passed through a MAP detector to detect which of the four mixture component is the source of that sample. These $10^5$ detections are then compared with their true sources to compute $C_d$ at each of time instant $T_3$. We found that this empirical total correct detections $ C_d $ is consistent with analytic $P_d$.
	\begin{figure}[t!] 
		\centering
		\includegraphics[width=.75\linewidth]{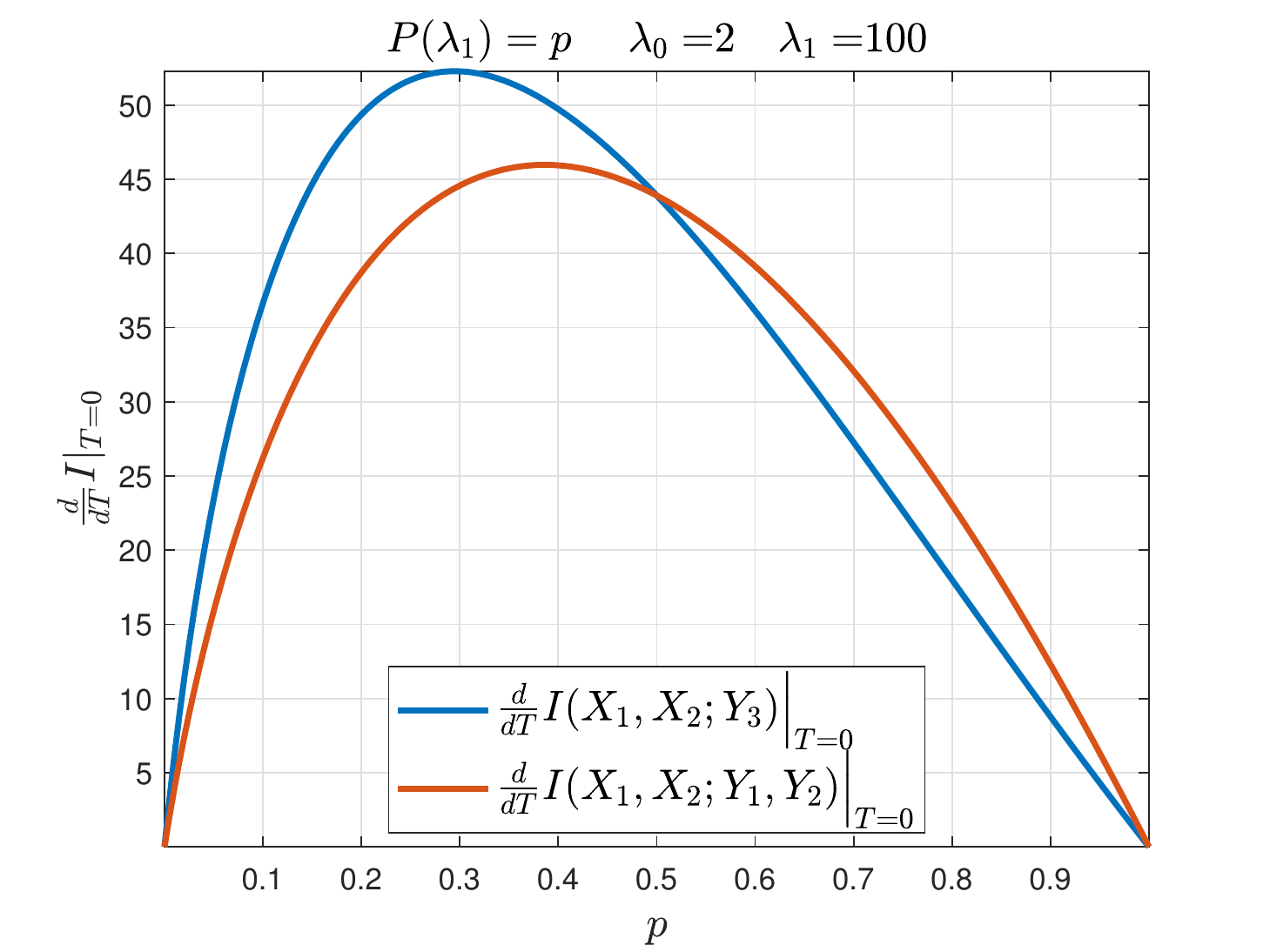}  
		\caption{Individual 2-target detection problem:  $\frac{d}{dT}I(X_1, X_2;Y_3) \Big|_{T=0}$ and $\frac{d}{dT}I(X_1, X_2;Y_1,Y_2) \Big|_{T=0}$ with $(T_1=\frac{T}{2}, T_2=\frac{T}{2})$ vs. $p$.	\todo[disable,inline]{FirstDerivative\_SumOfTargets.m} }
		\label{f07}
	\end{figure}
	The computing method for $I$ is described step-by-step in the flowcharts given in Fig.(\ref{L01}) and Fig.(\ref{L02}). The bounds on the entropy of Poisson variable are available in \cite{adell2010sharp}, we use them to validate that the computed Poisson entropies are within bounds.
	
	For an unconstrained problem of choosing individual sensing over jointing sensing for a given prior and fixed intensities for an infinitesimal or finite $T$; the first derivative of $I$ at $T=0$ given in \cite{guo2008mutual} (along with knowing the fact that $I$ is concave in $T$ \cite{atar2012mutual}) might be helpful. It is also helpful in verifying some of the numerical results of $I$ with analytical ones.
	\begin{figure}[t!] 
		\centering
		\includegraphics[width=.75\linewidth]{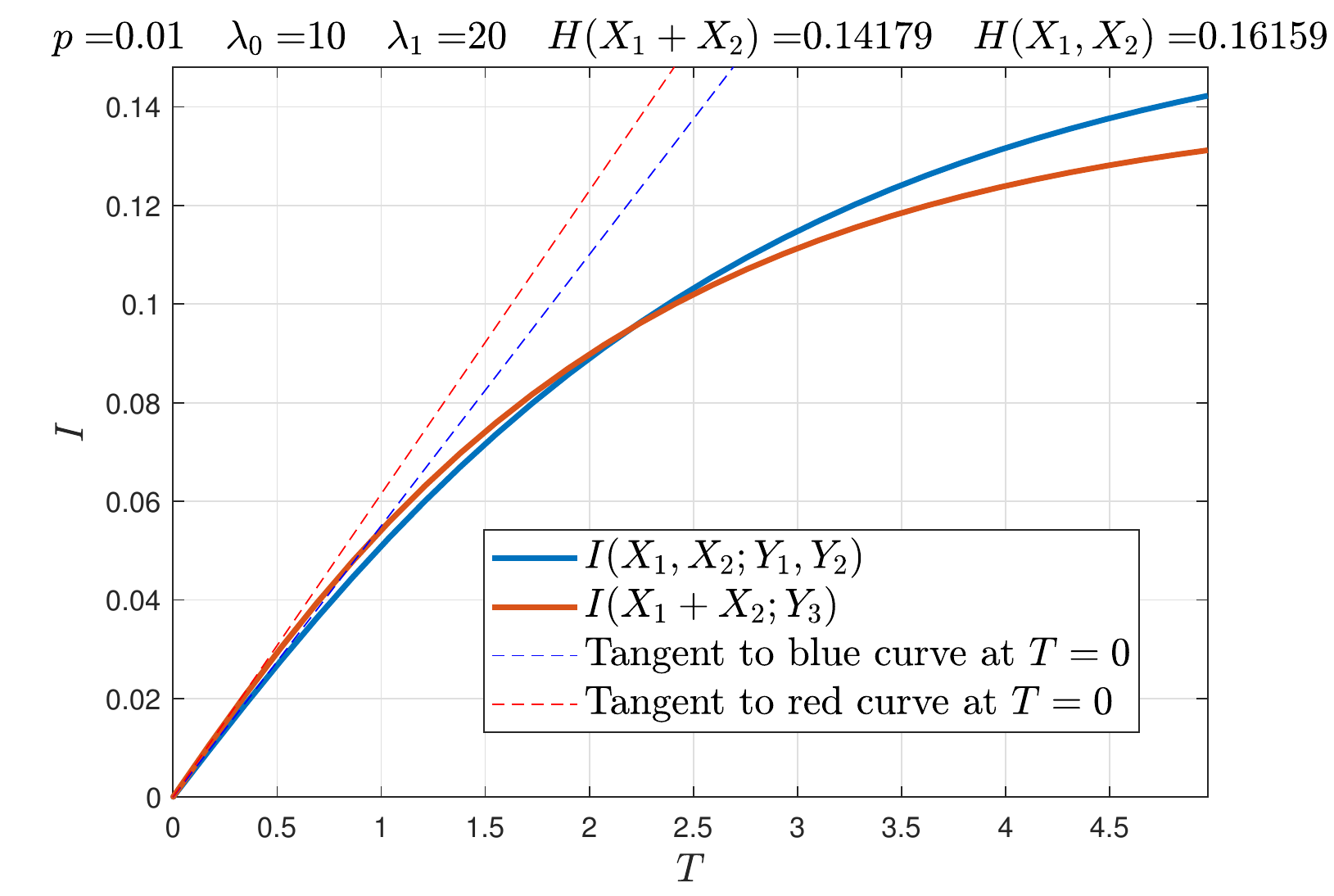}  
		\caption{Joint sensing and individual sensing vs. $T$. Note that joint sensing is better sensing method when $T < 2.25$. For $T>2.25$ individual sensing is better. 	\todo[disable,inline]{BinomialLimit\_I\_General\_N.m} }
		\label{f08}
	\end{figure}	
	In Fig.(\ref{f07}) we have $\frac{d}{dT}I(X_1, X_2;Y_3) \Big|_{T=0}$ and $\frac{d}{dT}I(X_1, X_2;Y_1,Y_2) \Big|_{T=0}$ vs. $p$ for $\lambda_0=2$ and $\lambda_1=100$. We can see that at $p=0.5$ the first derivatives of both the joint and individual sensing w.r.t time variable $T$ are same; this means that no matter what the given time is, individual sensing is never worse than the joint sensing at $p=0.5$. Whereas, for $p<0.5$ we need to further know what the given finite time $T$ is, to decide which of the two methods is better over the other for that given time $T$. As can be seen in Fig.(\ref{f08}) that the two $I$ curves cross each other, making one better over the other in two different time segments. Further note that in Fig. (\ref{f08}) the numerical derivative values of $I$ at $T=0$ are $0.055011541474291$ and $0.061490803759598$ for individual and joint sensing, respectively when $I$ is computed according to the algorithm defined in flowcharts. Now compare these numerical derivative values with analytically calculated ones $0.055011540931595$ and $0.061490807291534$; the two are accurate to eight decimal places, this validates that computations are accurate enough.
	
\begin{figure*}[ht]
		\begin{subfigure}{.49\textwidth}
			\includegraphics[width=\linewidth]{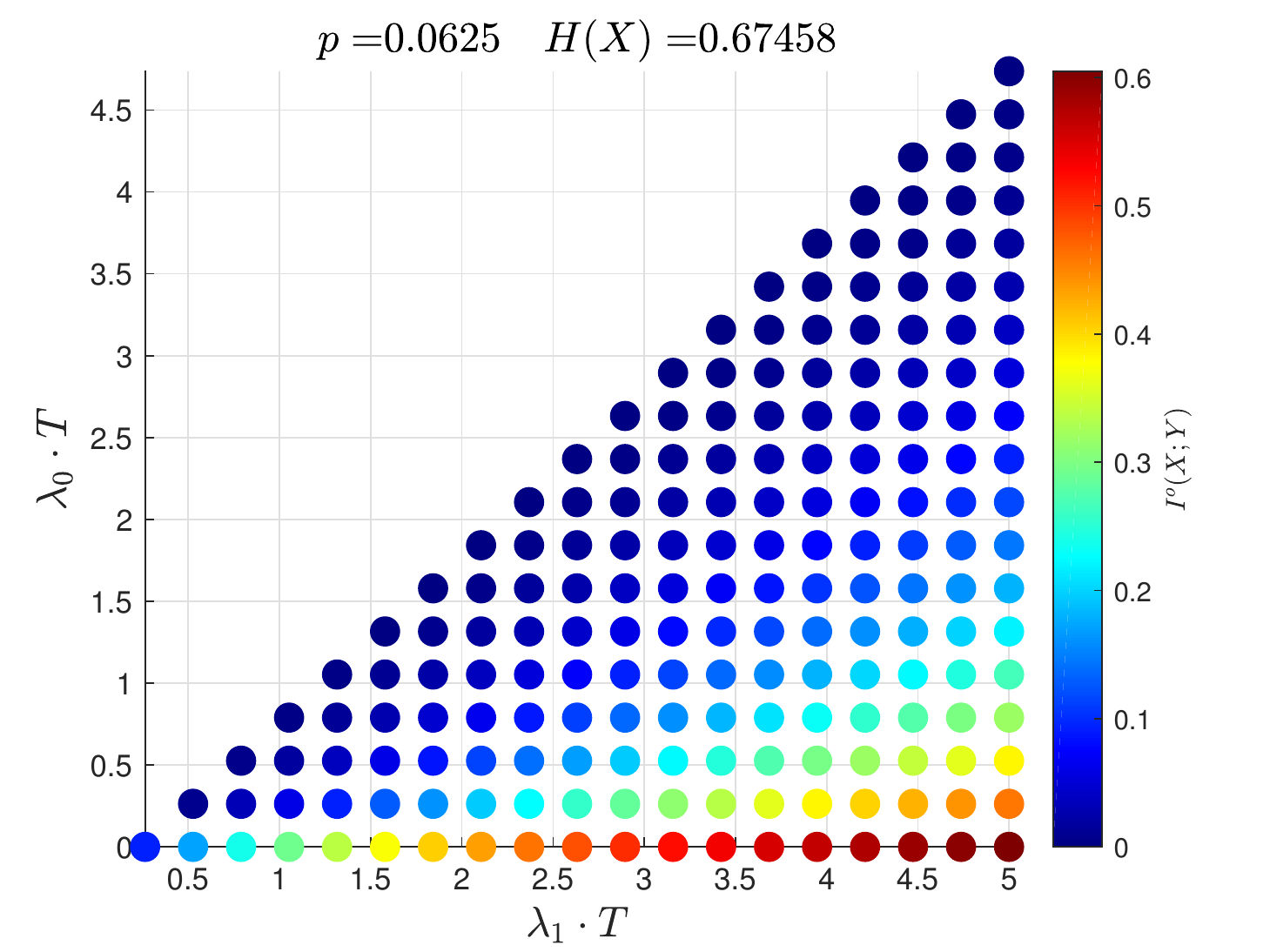}
			\caption{ }
			\label{fig6a}
		\end{subfigure} 
		\begin{subfigure}{.49\textwidth}
			\includegraphics[width=\linewidth]{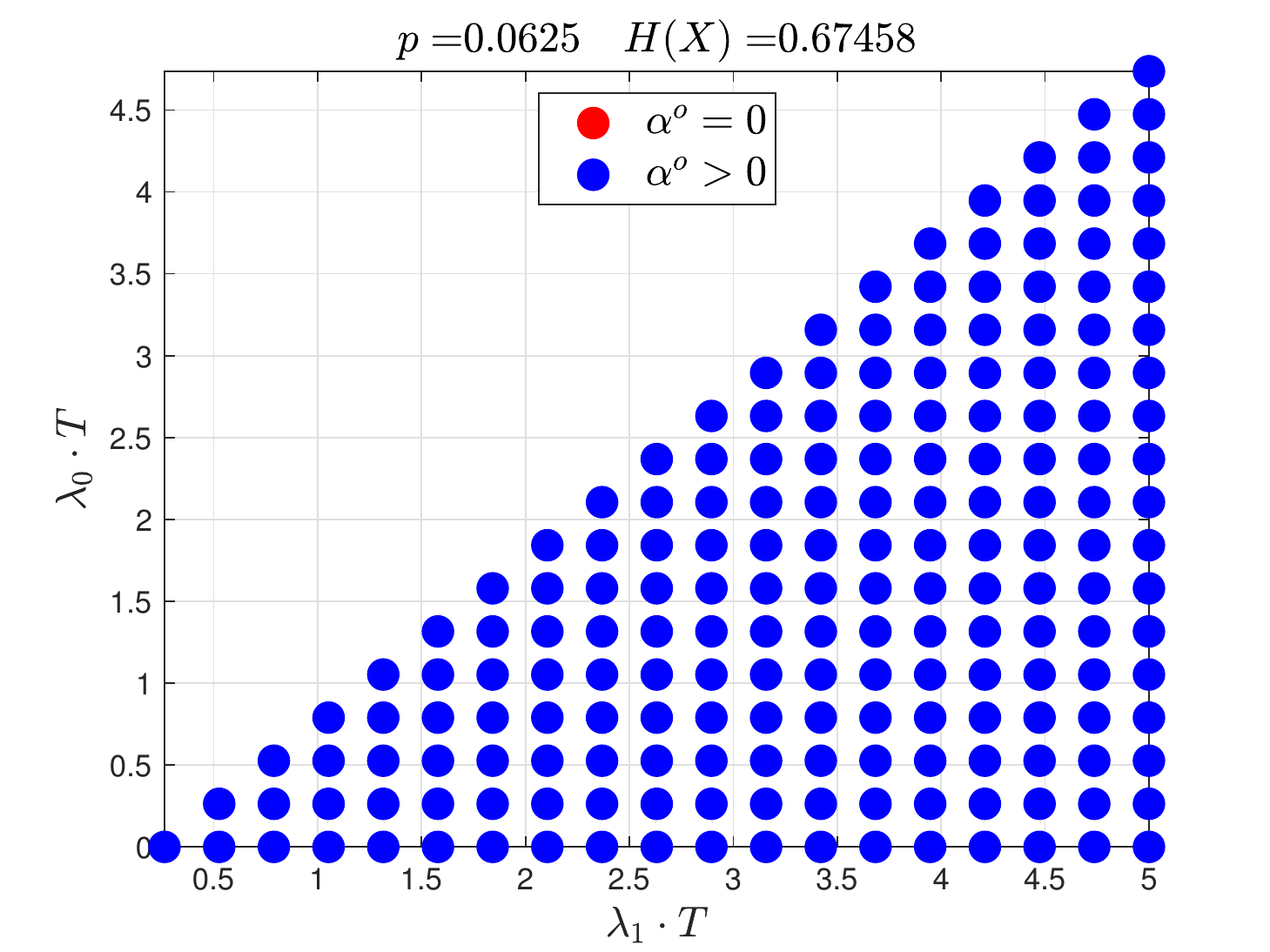}
			\caption{ }
			\label{fig6b}
		\end{subfigure} \\%
		\begin{subfigure}{.49\textwidth}
			\includegraphics[width=\linewidth]{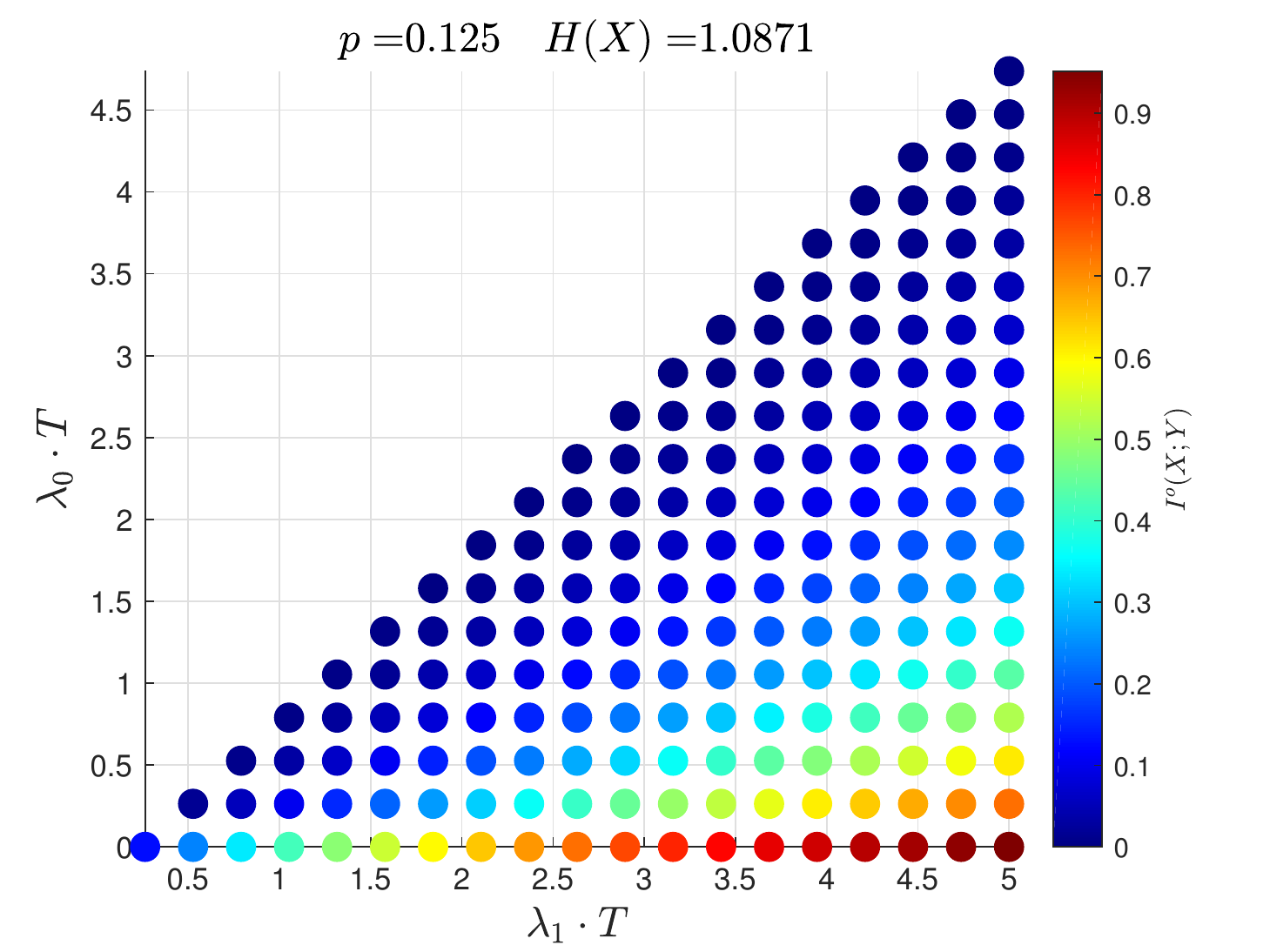}
			\caption{ }
			\label{fig6c}
		\end{subfigure} %
		\begin{subfigure}{.49\textwidth}
			\includegraphics[width=\linewidth]{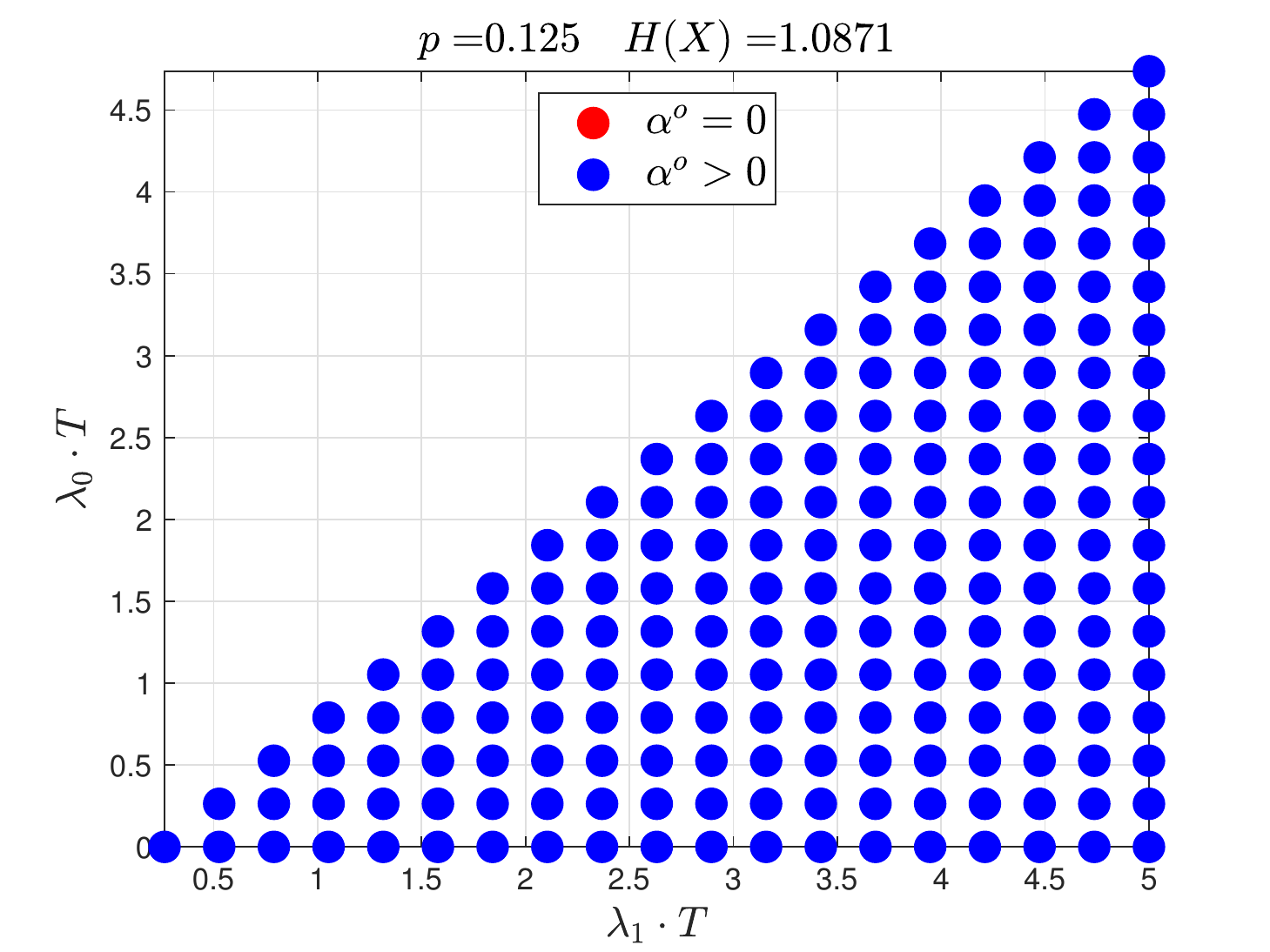}
			\caption{ }
			\label{fig6d}
		\end{subfigure}
		\begin{subfigure}{.49\textwidth}
			\includegraphics[width=\linewidth]{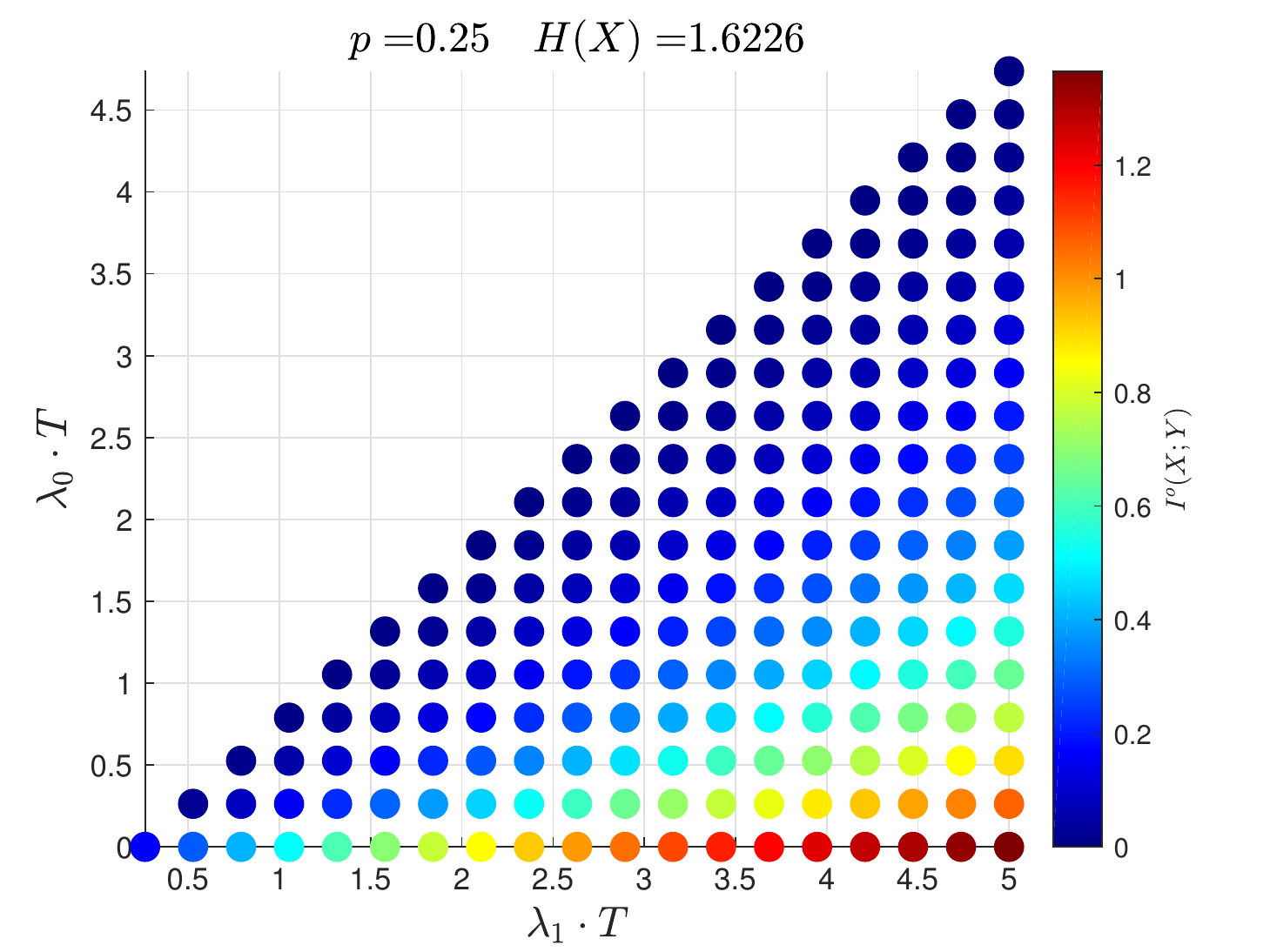}
			\caption{ }
			\label{fig6e}
		\end{subfigure} %
		\begin{subfigure}{.49\textwidth}
			\includegraphics[width=\linewidth]{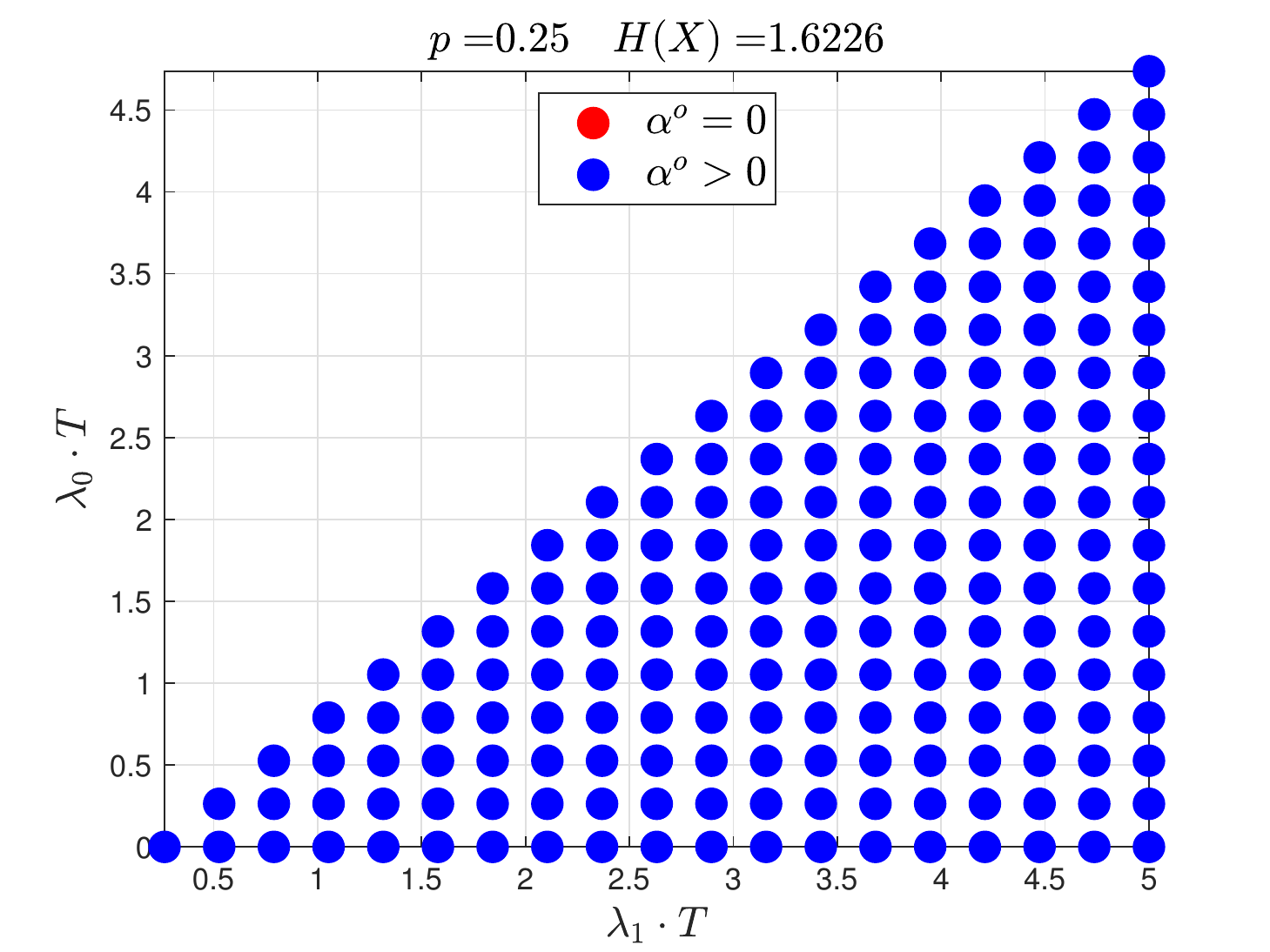}
			\caption{ }
			\label{fig6f}
		\end{subfigure} %
		\caption{Left: $ I^O(X;Y) $ vs. $ (\lambda_0 T,\lambda_1 T) $ in the region $\lambda_1 T > \lambda_0 T $, right: corresponding optimal argument parameter $\alpha^O$ vs. $ (\lambda_0 T,\lambda_1 T) $ for varying \emph{prior} probabilities $ p $. The search for each optimal argument $\alpha^O$ for any fixed: $(\lambda_0 T, \lambda_1 T)$ and $p$ is performed over the line $ (T_1,T_2,T_3) := (\frac{1-\alpha}{2},\frac{1-\alpha}{2},\alpha)$ where $0 \le \alpha \le 1$ and $T_1+T_2+T_3=1.$ \todo[disable,inline]{\texttt{\detokenize{Optimal_Alpha3_2.m}}
				\newline   	\texttt{\detokenize{my_MI_3.m}}                     }}
		\label{f6}
	\end{figure*}

\begin{figure*}[ht]
		\begin{subfigure}{.49\textwidth}
			\includegraphics[width=\linewidth]{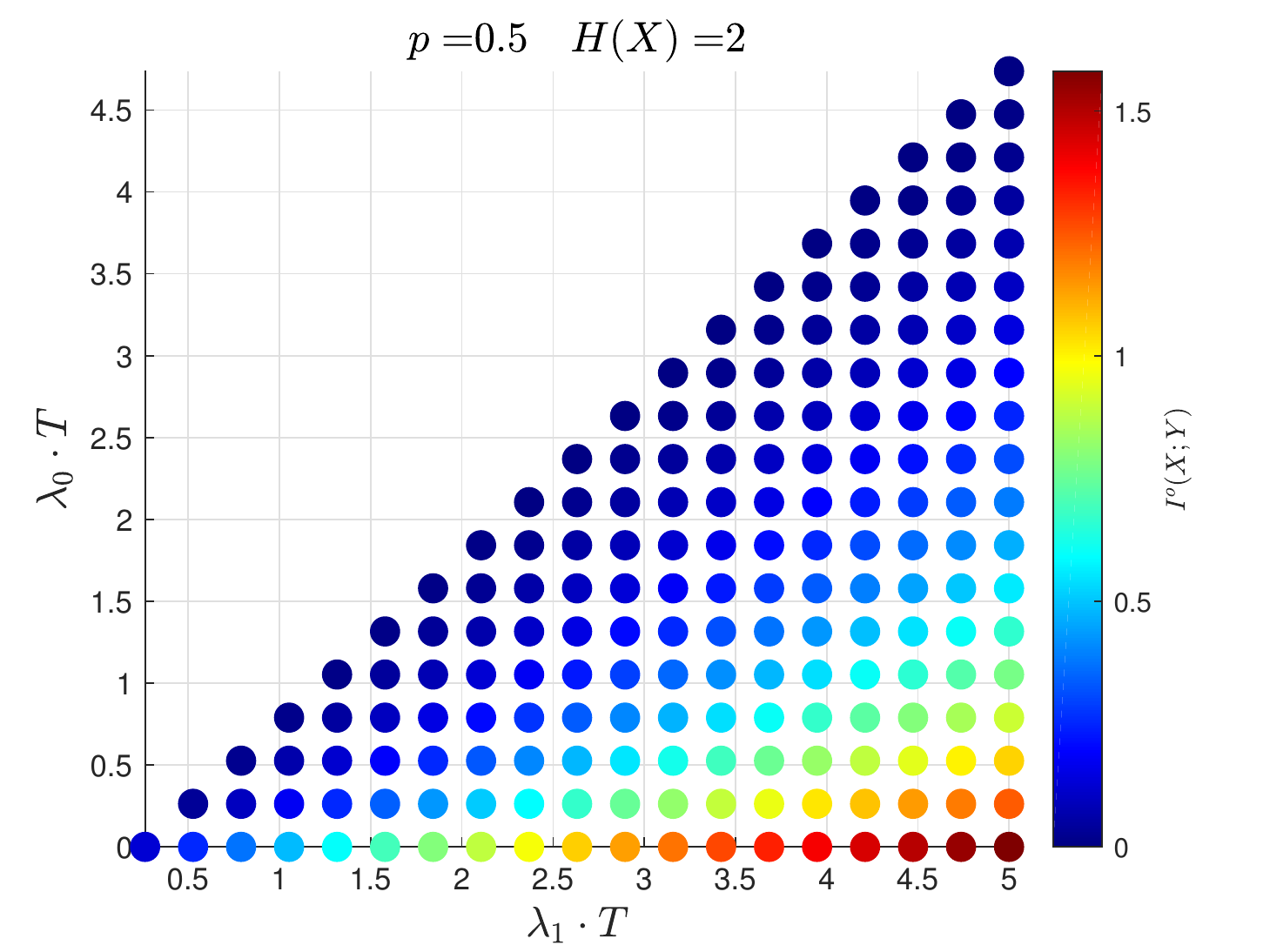}
			\caption{ }
			\label{fig7a}
		\end{subfigure} 
		\begin{subfigure}{.49\textwidth}
			\includegraphics[width=\linewidth]{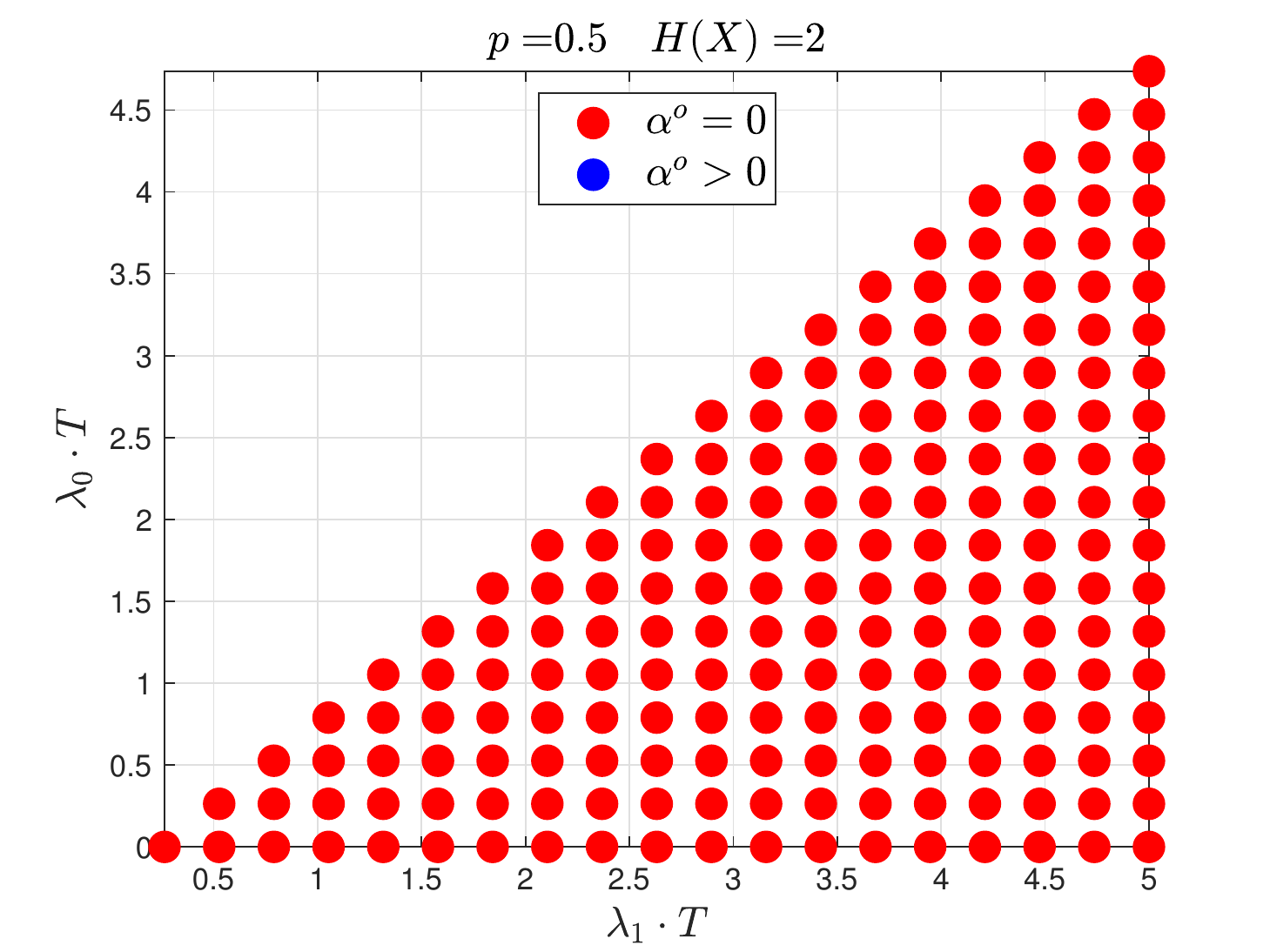}
			\caption{ }
			\label{fig7b}
		\end{subfigure} \\%
		\begin{subfigure}{.49\textwidth}
			\includegraphics[width=\linewidth]{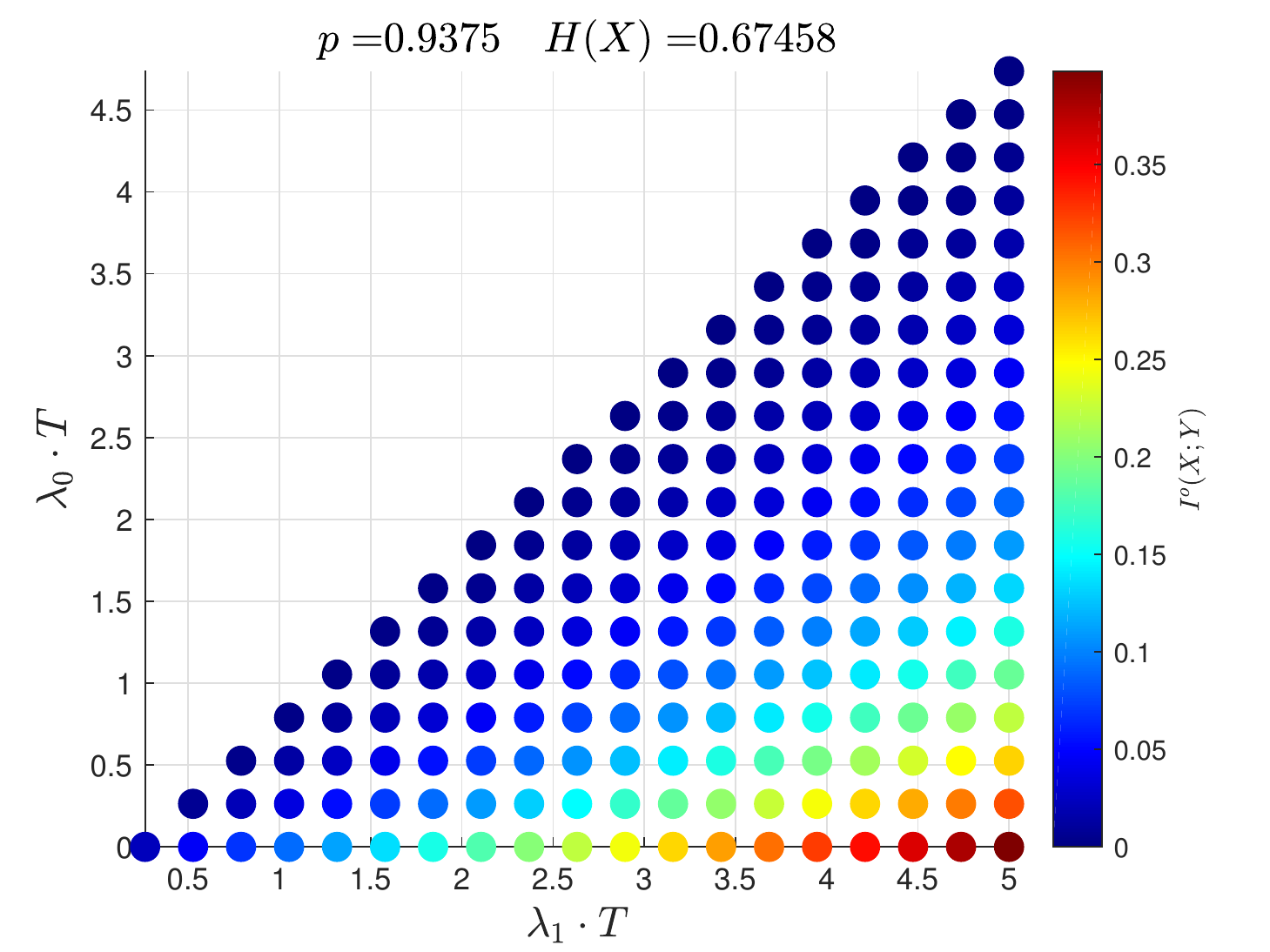}
			\caption{ }
			\label{fig7c}
		\end{subfigure} %
		\begin{subfigure}{.49\textwidth}
			\includegraphics[width=\linewidth]{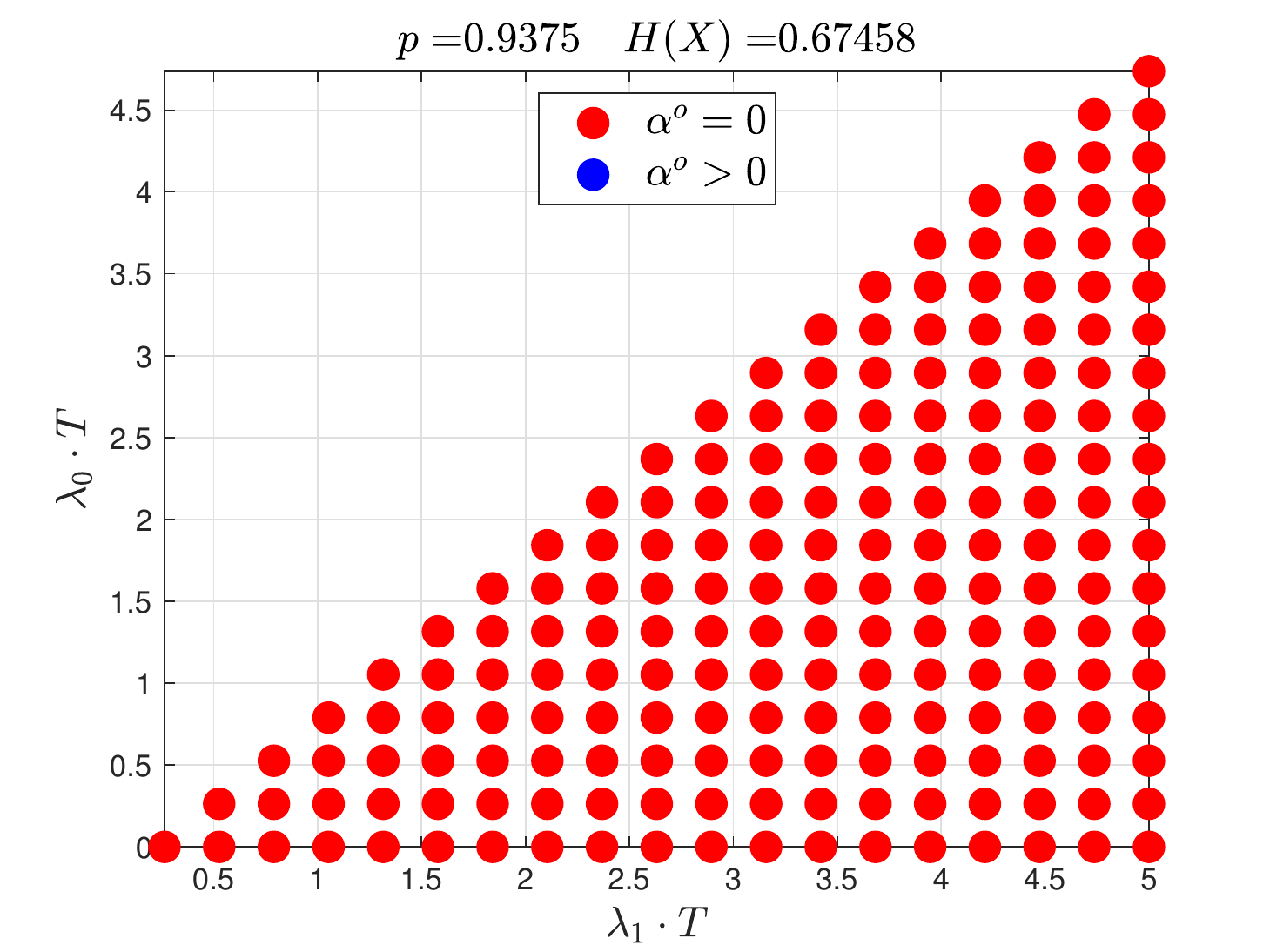}
			\caption{ }
			\label{fig7d}
		\end{subfigure}
		\begin{subfigure}{.49\textwidth}
			\includegraphics[width=\linewidth]{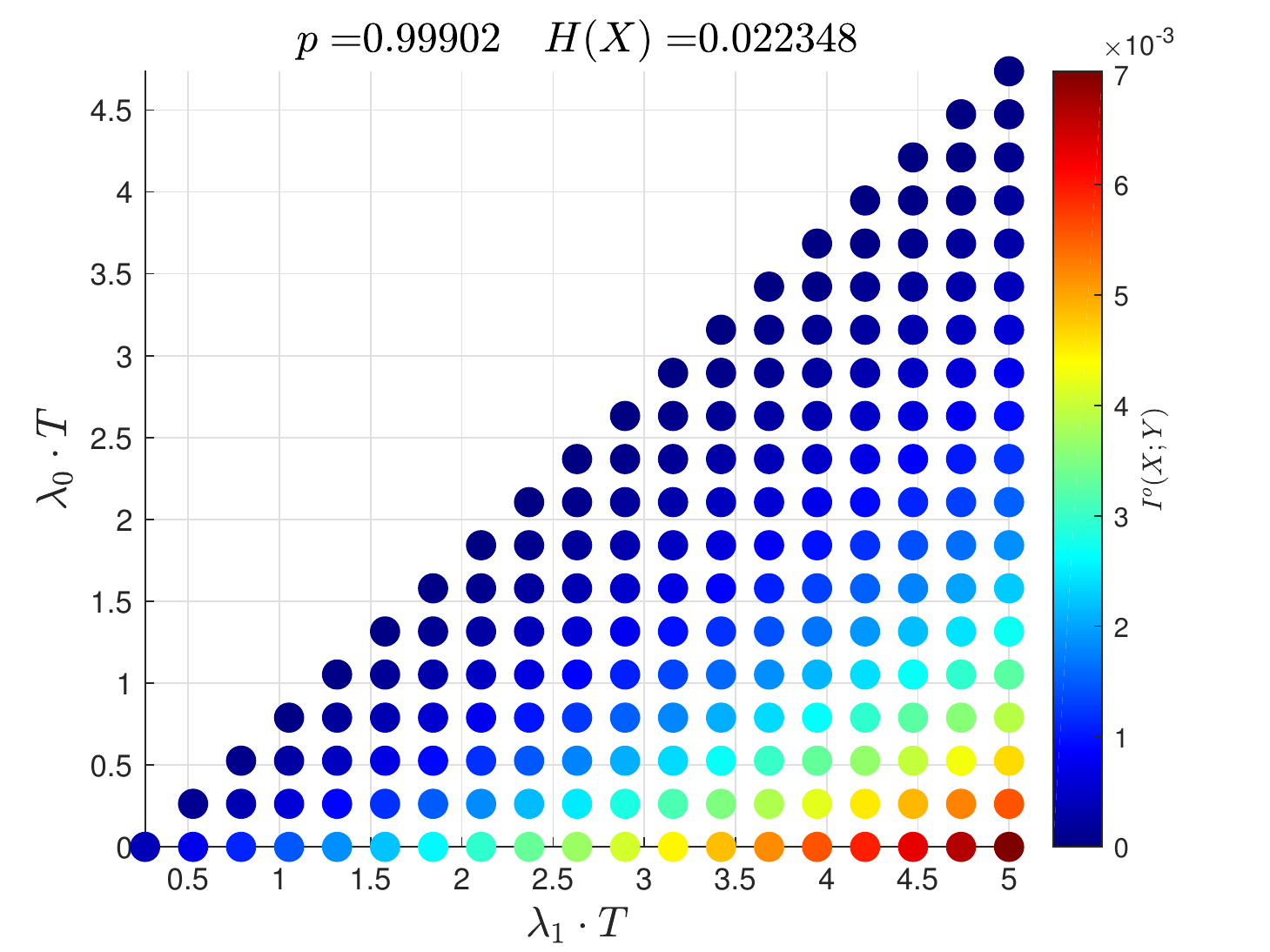}
			\caption{ }
			\label{fig7e}
		\end{subfigure} %
		\begin{subfigure}{.49\textwidth}
			\includegraphics[width=\linewidth]{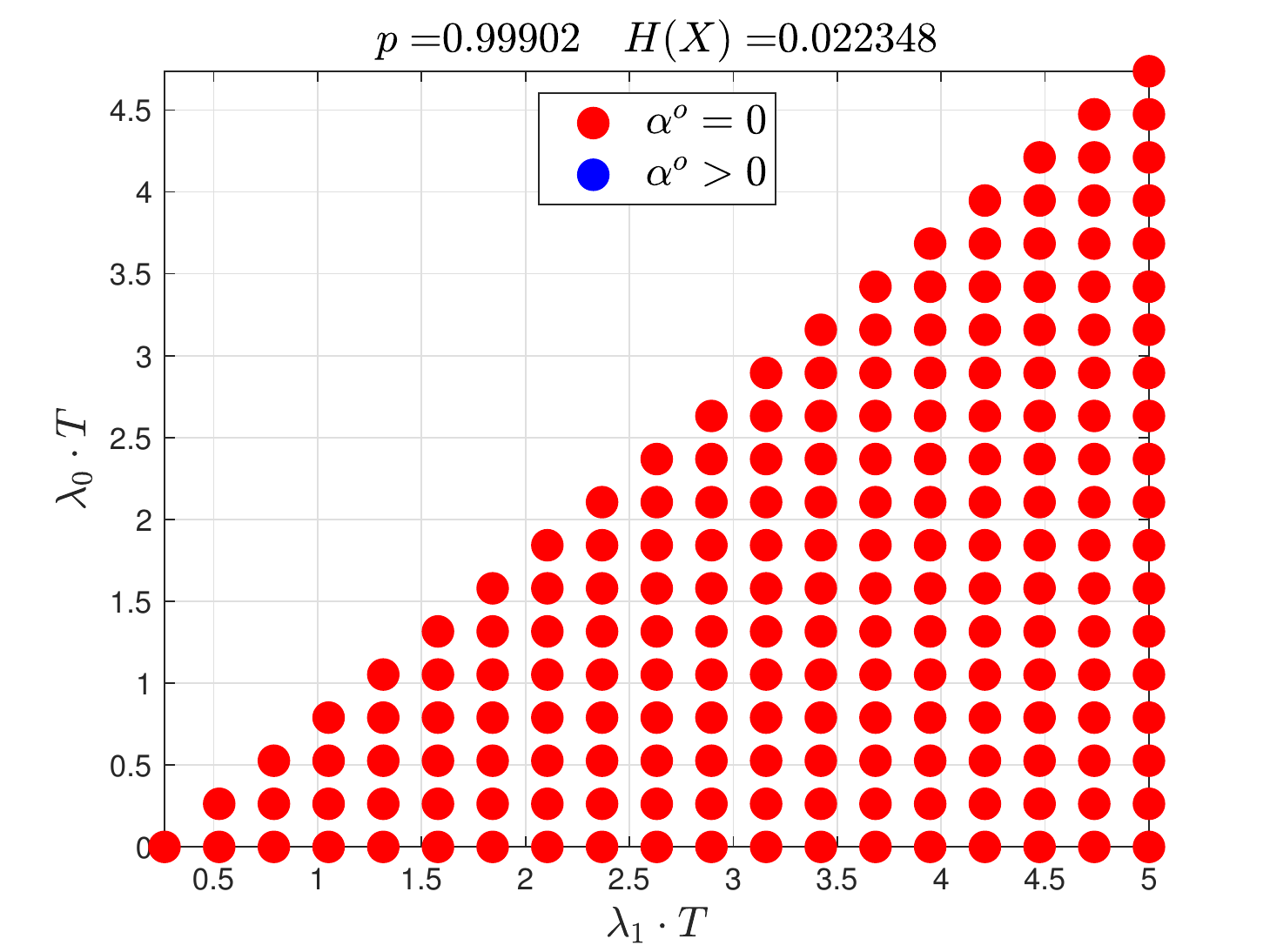}
			\caption{ }
			\label{fig7f}
		\end{subfigure} %
		\caption{Left: $ I^O(X;Y) $ vs. $ (\lambda_0 T,\lambda_1 T) $ in the region $\lambda_1 T > \lambda_0 T $, right: corresponding optimal argument parameter $\alpha^O$ vs. $ (\lambda_0 T,\lambda_1 T) $ for varying \emph{prior} probabilities $ p $. The search for each optimal argument $\alpha^O$ for any fixed: $(\lambda_0 T, \lambda_1 T)$ and $p$ is performed over the line $ (T_1,T_2,T_3) := (\frac{1-\alpha}{2},\frac{1-\alpha}{2},\alpha)$ where $0 \le \alpha \le 1$ and $T_1+T_2+T_3=1.$ \todo[disable,inline]{\texttt{\detokenize{Optimal_Alpha3_2.m}}
				\newline   	\texttt{\detokenize{my_MI_3.m}}                     }}
		\label{f7}
	\end{figure*}

	Fig. (\ref{f6}) and Fig. (\ref{f7}) illustrates the scatter plots, for a range of varying input intensities in constrained problem w.r.t $I$ metric. The scatter plots on the left hand side illustrates the optimal value of mutual information at any point $(\lambda_0 \cdot T, \lambda_1 \cdot T)$ where $ \lambda_0 \cdot T <  \lambda_1 \cdot T$ and $0 < \lambda_1 \cdot T \le 5 $ for a given prior $p$ and $T=1$. The corresponding maximizing argument is shown on the left hand side of the scatter plots. It is noticed that for prior $p<0.5$, farther the four components of multivariate mixture Poisson are from each other in terms of their intensities; more the optimal solution moves towards the individual sensing. In other words if the four pmfs (under hypothesis testing) gets closer, more the optimal sensing relies on joint sensing for prior $p<0.5$. If the prior $p \ge 0.5$ then irrespective of the other parameter values in the model, individual sensing is the optimal strategy from $I$ metric.
	\section{Conclusion} \label{con}
	In this work we have addressed a simple abstract sensor scheduling problem in which two Poisson sources are observed either individually or jointly with a switching mechanism and a single counter.  The scheduling problem is to choose the times associated with the different switch configurations given a total time constraint.
	
	For our specific problem we were able to solve the optimization problem using computational methods, primarily because of the relatively small number of free parameters.  For the case in which mutual information is the cost function, we observed but were unable to prove that the cost function is concave in the switch times; if true this reduces the optimization problem to a one-dimensional search.  We observed that the two different metrics of mutual information and probability of correct detection did not always lead to the same scheduling solution.  A second interesting result was that, under the mutual information metric, when the prior $p$ on the sources was greater than 0.5, individual sensing was always the best that could be done.
	
	In a certain sense our computational results are not particularly interesting.  In most cases the cost functions are quite flat and there were no dramatic variation in mutual information or probability of correct detection across the space of possible switch configuration times.  A communications engineer faced with the problem exactly as we have posed it would be hard-pressed to justify doing anything other than the simplest and most obvious solution, namely taking the available time and dividing it equally between individual observation of the two sources.
	
	However, the fact that there is \emph{some} variation in performance based in intelligent choice of the switch times leaves open the possibility of the value of investigations like ours for more complex problems.  Several extensions immediately come to mind: more sources (perhaps hundreds or thousands, not two), unknown Poisson rates, different unknown or nonexistent priors, different source distributions.  Unfortunately, the jump to a larger number of sources brings with it a combinatorial explosion which would rule out the brute-force optimization techniques used in this paper in very short order.  Alternate switching strategies might include an adaptive one in which the switching time and configurations are determined online or in real time depending on the observations up to the current time. We recommend and are interested in continuing investigations along these lines.		 
	
	\appendices
	\section{Expression of mutual information $ I(X_1,X_2;Y_1,Y_2,Y_3) $ } \label{mixy}
	\begin{IEEEeqnarray*}{rCl}
		&&I(X;Y)  =- \sum\limits_{y_1=0}^{\infty} \: \sum\limits_{y_2=0}^{\infty} \: \sum\limits_{y_3=0}^{\infty} \: \Bigg[ \Bigg( (1-p)^2 \cdot  \operatorname{Poiss}(y_1; \lambda_0 T_1) \cdot  \operatorname{Poiss}(y_2; \lambda_0 T_2) \cdot  \\ && \> \operatorname{Poiss}(y_3; 2\lambda_0 T_3)  + p(1-p) \cdot\operatorname{Poiss}(y_1; \lambda_0 T_1) \cdot   \operatorname{Poiss}(y_2; \lambda_1 T_2) \cdot  \operatorname{Poiss}(y_3; (\lambda_0+\lambda_1) T_3)   + \\ && \> p(1-p) \cdot  \operatorname{Poiss}(y_1; \lambda_1 T_1) \cdot \operatorname{Poiss}(y_2; \lambda_0 T_2) \cdot  \operatorname{Poiss}(y_3; (\lambda_1+\lambda_0) T_3)   + p^2 \cdot \operatorname{Poiss}(y_1; \lambda_1 T_1) \cdot  \\ && \>  \operatorname{Poiss}(y_2; \lambda_1 T_2) \cdot  \operatorname{Poiss}(y_3; 2 \lambda_1 T_3) \Bigg)  \cdot  \Bigg( \operatorname{log}_2[(1-p)^2 \cdot  \operatorname{Poiss}(y_1; \lambda_0 T_1) \cdot \operatorname{Poiss}(y_2; \lambda_0 T_2) \cdot   \\ && \> \operatorname{Poiss}(y_3; 2\lambda_0 T_3)  + p(1-p) \cdot \operatorname{Poiss}(y_1; \lambda_0 T_1) \cdot  \operatorname{Poiss}(y_2; \lambda_1 T_2) \cdot  \operatorname{Poiss}(y_3; (\lambda_0+\lambda_1) T_3)   +  \\ && \>  p(1-p) \cdot \operatorname{Poiss}(y_1; \lambda_1 T_1) \cdot \operatorname{Poiss}(y_2; \lambda_0 T_2) \cdot  \operatorname{Poiss}(y_3; (\lambda_1+\lambda_0) T_3)   + \\ && \> p^2 \cdot \operatorname{Poiss}(y_1; \lambda_1 T_1) \cdot \operatorname{Poiss}(y_2; \lambda_1 T_2) \cdot  \operatorname{Poiss}(y_3; 2 \lambda_1 T_3)] \Bigg) \Bigg] + \\ && \> \Bigg[ \Bigg( (1-p)^2 \cdot \sum\limits_{y_1=0}^{\infty} \sum\limits_{y_2=0}^{\infty} \sum\limits_{y_3=0}^{\infty} (\operatorname{Poiss}(y_1; \lambda_0 T_1) \cdot \operatorname{Poiss}(y_2; \lambda_0 T_2) \cdot \operatorname{Poiss}(y_3; 2\lambda_0 T_3))\cdot \\ && \>  \operatorname{log}_2[\operatorname{Poiss}(y_1; \lambda_0 T_1) \cdot \operatorname{Poiss}(y_2; \lambda_0 T_2) \cdot  \operatorname{Poiss}(y_3; 2\lambda_0 T_3)] \Bigg) + \\ && \> \Bigg( p(1-p) \cdot \sum\limits_{y_1=0}^{\infty} \sum\limits_{y_2=0}^{\infty} \sum\limits_{y_3=0}^{\infty} (\operatorname{Poiss}(y_1; \lambda_0 T_1) \cdot \operatorname{Poiss}(y_2; \lambda_1 T_2) \cdot \operatorname{Poiss}(y_3; (\lambda_0 + \lambda_1) T_3)\cdot \\ && \> \operatorname{log}_2[\operatorname{Poiss}(y_1; \lambda_0 T_1) \cdot \operatorname{Poiss}(y_2; \lambda_1 T_2) \cdot   \operatorname{Poiss}(y_3; (\lambda_0 + \lambda_1) T_3)] \Bigg) + \Bigg( p(1-p) \cdot \\ && \> \sum\limits_{y_1=0}^{\infty} \sum\limits_{y_2=0}^{\infty} \sum\limits_{y_3=0}^{\infty}  (\operatorname{Poiss}(y_1; \lambda_1 T_1)  \cdot \operatorname{Poiss}(y_2; \lambda_0 T_2) \cdot  \operatorname{Poiss}(y_3; (\lambda_1 + \lambda_0) T_3)\cdot \\ && \> \operatorname{log}_2[\operatorname{Poiss}(y_1; \lambda_1 T_1) \cdot   \operatorname{Poiss}(y_2; \lambda_0 T_2) \cdot\operatorname{Poiss}(y_3; (\lambda_1 + \lambda_0) T_3)] \Bigg)  + \\ && \> \Bigg( p^2 \cdot \sum\limits_{y_1=0}^{\infty} \sum\limits_{y_2=0}^{\infty} \sum\limits_{y_3=0}^{\infty} (\operatorname{Poiss}(y_1; \lambda_1 T_1) \cdot   \operatorname{Poiss}(y_2; \lambda_1 T_2) \cdot\operatorname{Poiss}(y_3; 2 \lambda_1 T_3)\cdot \\ && \>  \operatorname{log}_2[(\operatorname{Poiss}(y_1; \lambda_1 T_1) \cdot \operatorname{Poiss}(y_2; \lambda_1 T_2) \cdot  \operatorname{Poiss}(y_3; 2 \lambda_1 T_3)] \Bigg)               \Bigg]. \label{e14} 
	\end{IEEEeqnarray*}
	\clearpage
	\section{Expression of Probability of total correct detections $ P_d $ } \label{pd}
	\begin{IEEEeqnarray*}{rCl}
		&&P_d =\\ && \> \sum\limits_{(y_1,y_2,y_3)}^{}f_{00} : \{ (f_{00} >= f_{01}) \: \& \: (f_{00} >= f_{10}) \: \& \: (f_{00} >= f_{11})  \} + \\ && \> \sum\limits_{(y_1,y_2,y_3)}^{}f_{01} : \{ (f_{01} > f_{00}) \: \& \: (f_{01} >= f_{10}) \: \& \: (f_{01} >= f_{11}))  \} +  \\ && \> \sum\limits_{(y_1,y_2,y_3)}^{}f_{10} : \{ (f_{10} > f_{00}) \: \& \: (f_{10} > f_{01}) \: \& \: (f_{10} >= f_{11}) \} +  \\ && \>  \sum\limits_{(y_1,y_2,y_3)}^{}f_{11} : \{ (f_{11} > f_{00}) \: \& \: (f_{11} > f_{01}) \: \& \: (f_{11} > f_{10})  \}.
	\end{IEEEeqnarray*}
	Where
	\begin{IEEEeqnarray*}{rCl}
		f_{00}& \equiv &  (1-p)^2 \cdot  \operatorname{Poiss}(y_1; \lambda_0 T_1) \cdot \operatorname{Poiss}(y_2; \lambda_0 T_2)  \cdot   \operatorname{Poiss}(y_3; 2\lambda_0 T_3), \\
		f_{01} & \equiv &  p(1-p) \cdot  \operatorname{Poiss}(y_1; \lambda_0 T_1) \cdot \operatorname{Poiss}(y_2; \lambda_1 T_2)  \cdot  \operatorname{Poiss}(y_3; (\lambda_0 + \lambda_1) T_3), \\
		f_{10}& \equiv &  p(1-p) \cdot  \operatorname{Poiss}(y_1; \lambda_1 T_1) \cdot \operatorname{Poiss}(y_2; \lambda_0 T_2)  \cdot  \operatorname{Poiss}(y_3; (\lambda_1 + \lambda_0) T_3), \\
		f_{11}& \equiv &  p^2 \cdot  \operatorname{Poiss}(y_1; \lambda_1 T_1) \cdot \operatorname{Poiss}(y_2; \lambda_1 T_2)  \cdot  \operatorname{Poiss}(y_3; 2\lambda_1 T_3), \\  \& &  : & \mathrm{logical \: AND \: operation}
	\end{IEEEeqnarray*}
	\section{Numerical approximation of the Poisson pmf and mutual information}\label{mi}
	\begin{IEEEeqnarray*}{rCl}
		\operatorname{Poiss}(x; \lambda) \approx \left\{
		\begin{array}{ll}
			\frac{e^{\lambda}\lambda^x}{x!} & \quad  x   \le x_c  \\ 
			0 & \quad x> x_c  \\             
		\end{array}
		\right. \label{v8} \IEEEyesnumber
	\end{IEEEeqnarray*}
	where $ x_c $ is such that cummulative distribution function of Poisson random variable at $ x_c $ is approximately equal to $1.$ i.e. $\operatorname{CDFPoiss}(x_c;\lambda) \approx 1 $. When represented in double precision floating-point arithmetic in IEEE 754 standard; this means that numerical values of $\operatorname{CDFPoiss}(x_c;\lambda) $ and   $\operatorname{CDFPoiss}(x_c+1;\lambda) $ are identical and equal to numerical one.
	
	The infinite summations in mutual information expression given in appendix \ref{mixy} are all truncated from 0 to $ x_c=\operatorname{CDFPoiss}^{-1}(1-\epsilon;(\lambda_1+\lambda_1)T) $ for a given value of $\lambda_1$ and $ T $. $ \operatorname{CDFPoiss}^{-1} $ is inverse Poisson cdf and $ \epsilon $ is, the machine precision number, $2^{-53} $ \cite{trefethen1997numerical}.
	\section{Code description}
	\label{flowchart}
\begin{figure}[!h] 
		\centering
		\includegraphics[width=0.5\linewidth,trim={0.5inch 1.75inch 4.75inch 0.5inch},clip=true]{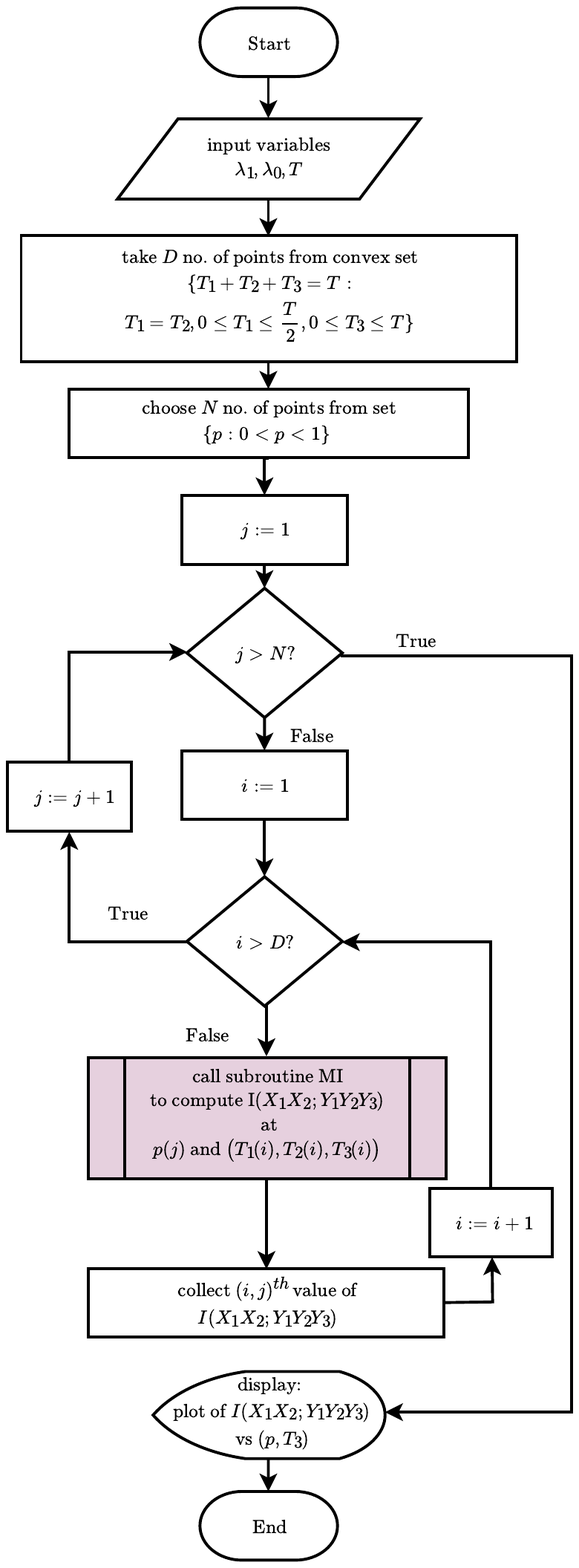}
		\caption{Flowchart 1: Algorithm description for computation of $I$.
			\todo[disable,inline]{Flowchart Diagram.xml} }
		\label{L01}
	\end{figure}

\begin{figure}[!h] 
		\centering
		\includegraphics[width=0.8\linewidth,trim={3.7inch 2.5inch 1.25inch 0.5inch},clip=true]{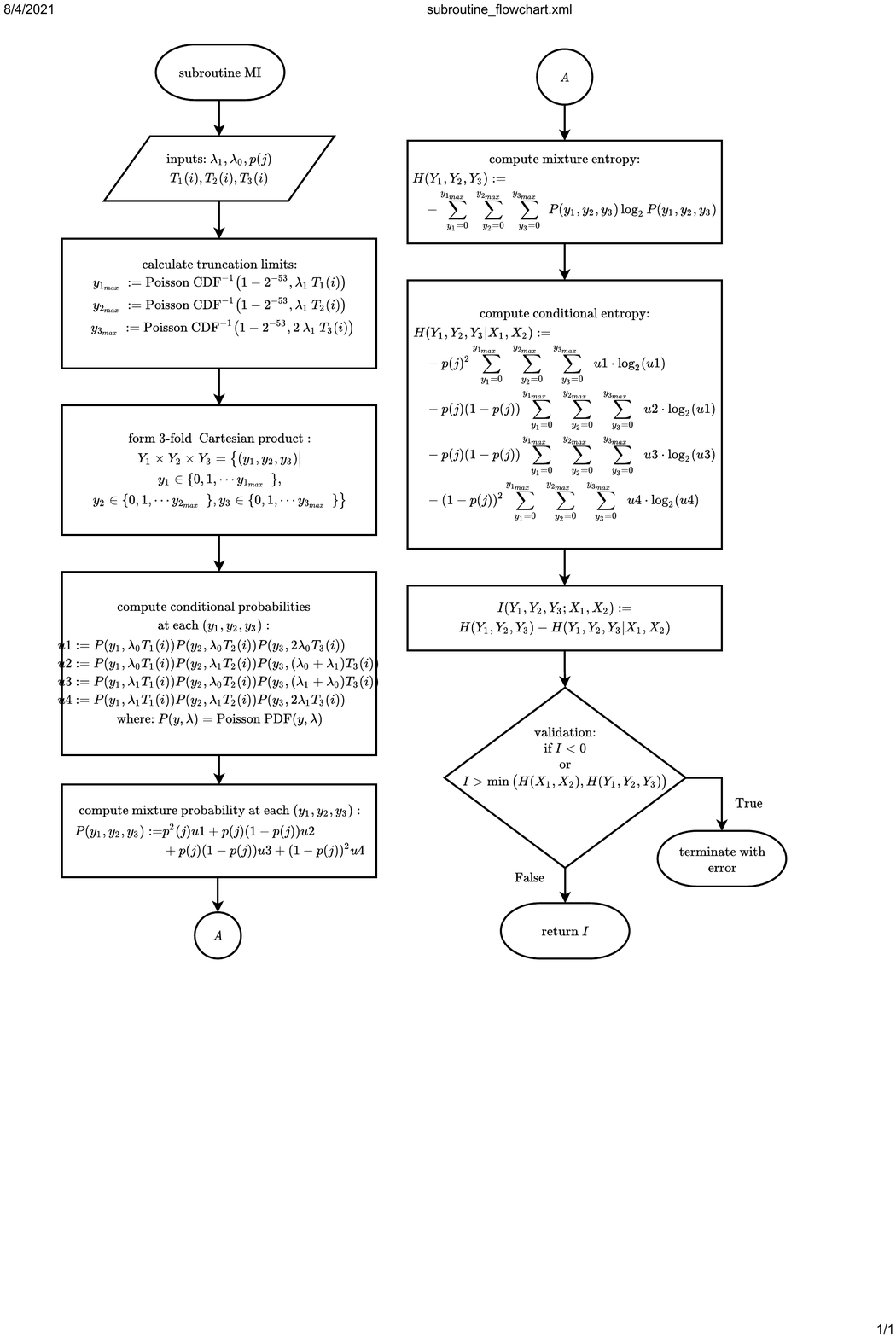}
		\caption{Flowchart 2: subroutine for computation of $I$.
			\todo[disable,inline]{subroutine\_flowchart.xml} }
		\label{L02}
	\end{figure}

	Since computations for $I$ can be done in parallel, we may exploit this fact by vectorization of the algorithm and then evaulating $I$ at grid points in parallel as defined in flowchart in Fig.(\ref{L01}) and Fig.(\ref{L02}) for fast computational throughput. However, it must be noted that this way of vectorized computations by first performing the truncation of Poisson distribution and then forming the grid of points, for further calculation of $I$, is not effective for large values of Poisson intensities involved in the problem. This is because the grid size increases cubely w.r.t truncated support of involved pmfs. So, for large values of Poisson intensities, say $\lambda >> 100$, we need to resort to Monte Carlo method. 
	
	For calculation of $P_d$, the same computed conditional Poisson pmfs can be used for all the comparisons that are required in conditional expression given in Appendix \ref{pd}.
	\twocolumn
	\begin{figure}[!h] 
		\centering
		\includegraphics[width=0.8\linewidth,trim={0.8inch 2.5inch 4.9inch 0.5inch},clip=true]{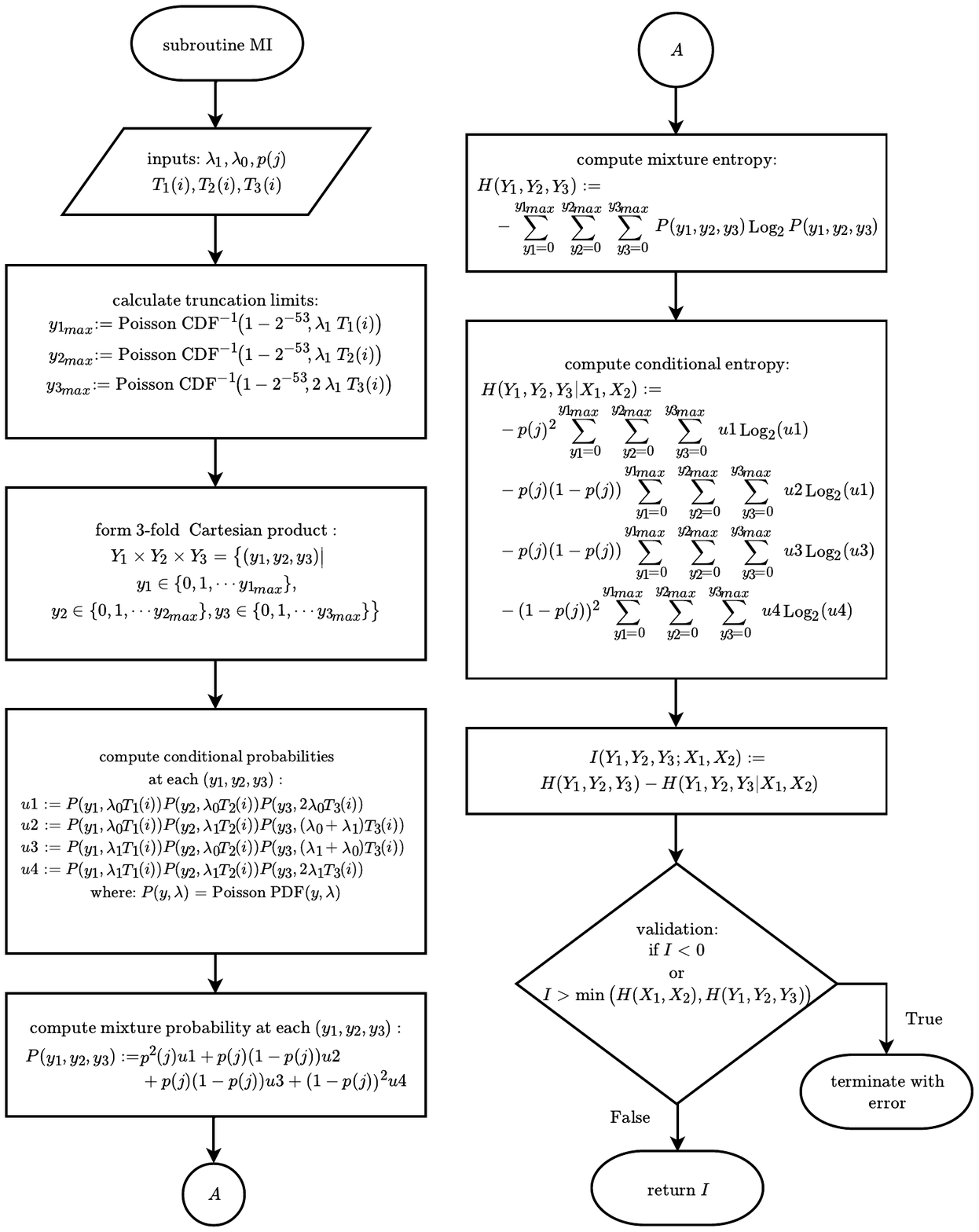}
	\end{figure}
	\onecolumn
	\begin{figure}[t]
		\centering
				\begin{subfigure}[b]{.49\textwidth}
					\centering
					\includegraphics[width=1\linewidth,height=50mm]{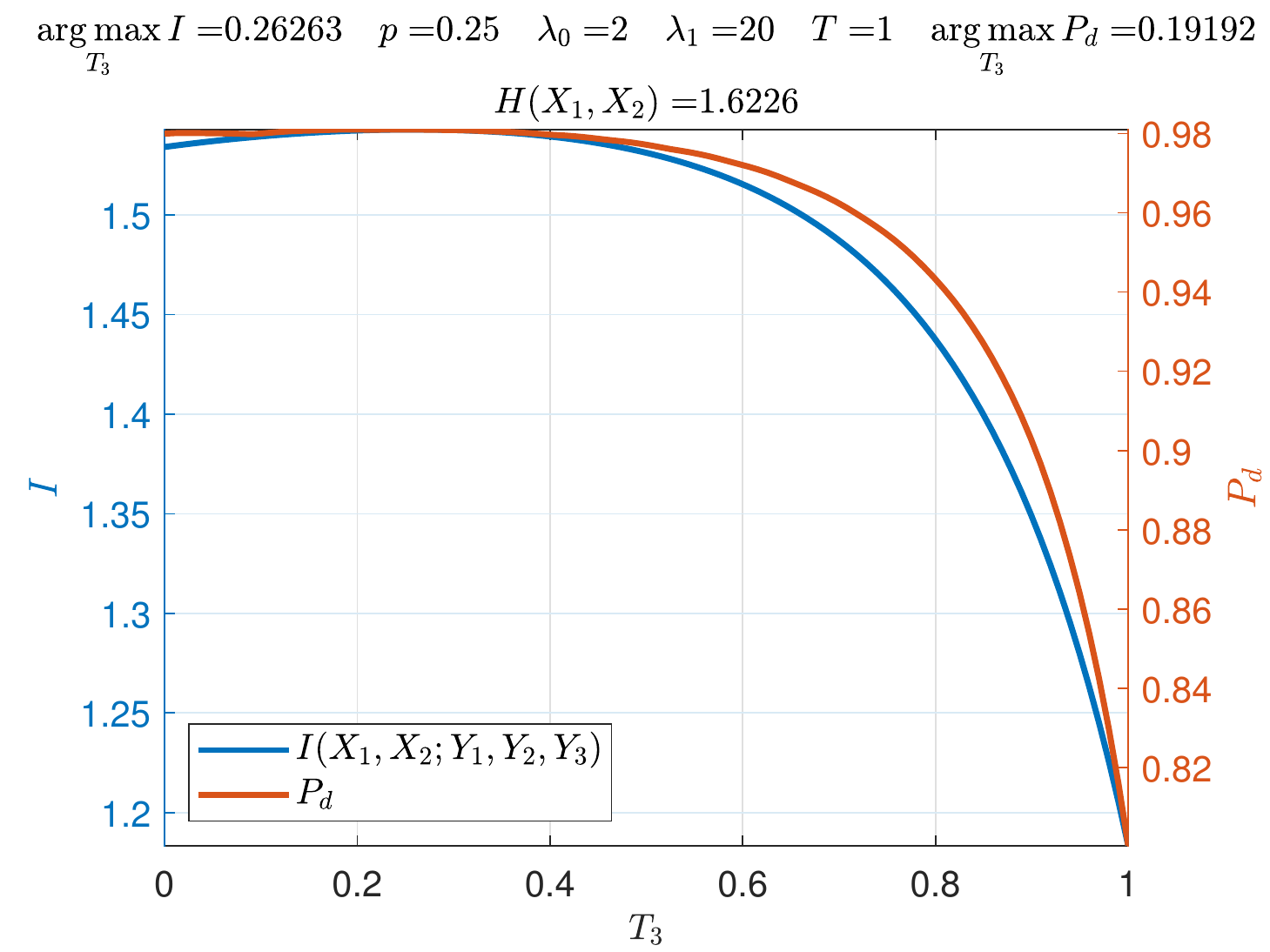}
					\caption{}
					\label{fig3a}
		\end{subfigure}
		\begin{subfigure}[b]{.49\textwidth}
			\centering
			\includegraphics[width=1\linewidth,height=50mm]{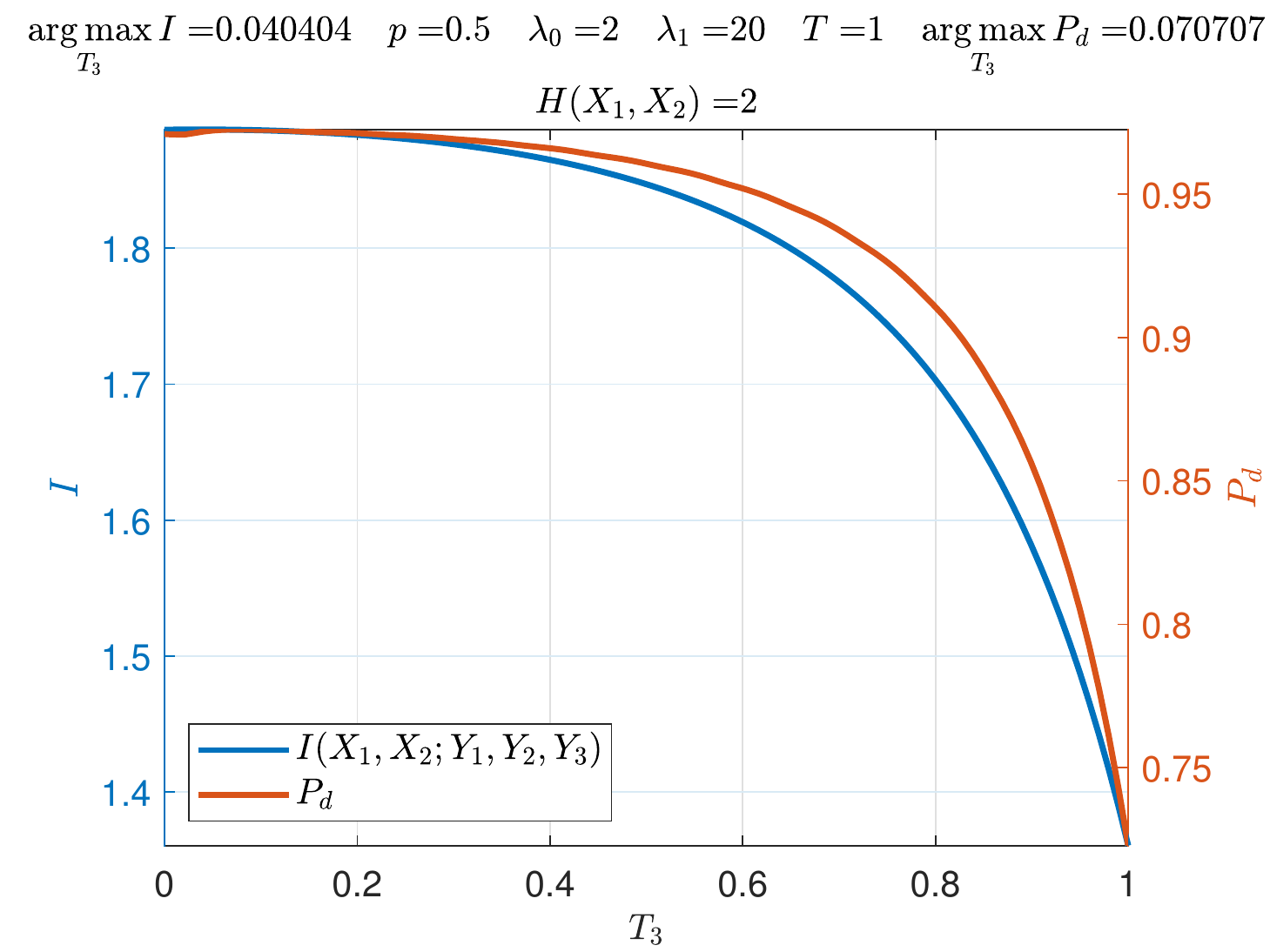}
			\caption{}
			\label{fig3b}
		\end{subfigure} %
		\begin{subfigure}[b]{.49\textwidth}
			\centering
			\includegraphics[width=1\linewidth,height=50mm]{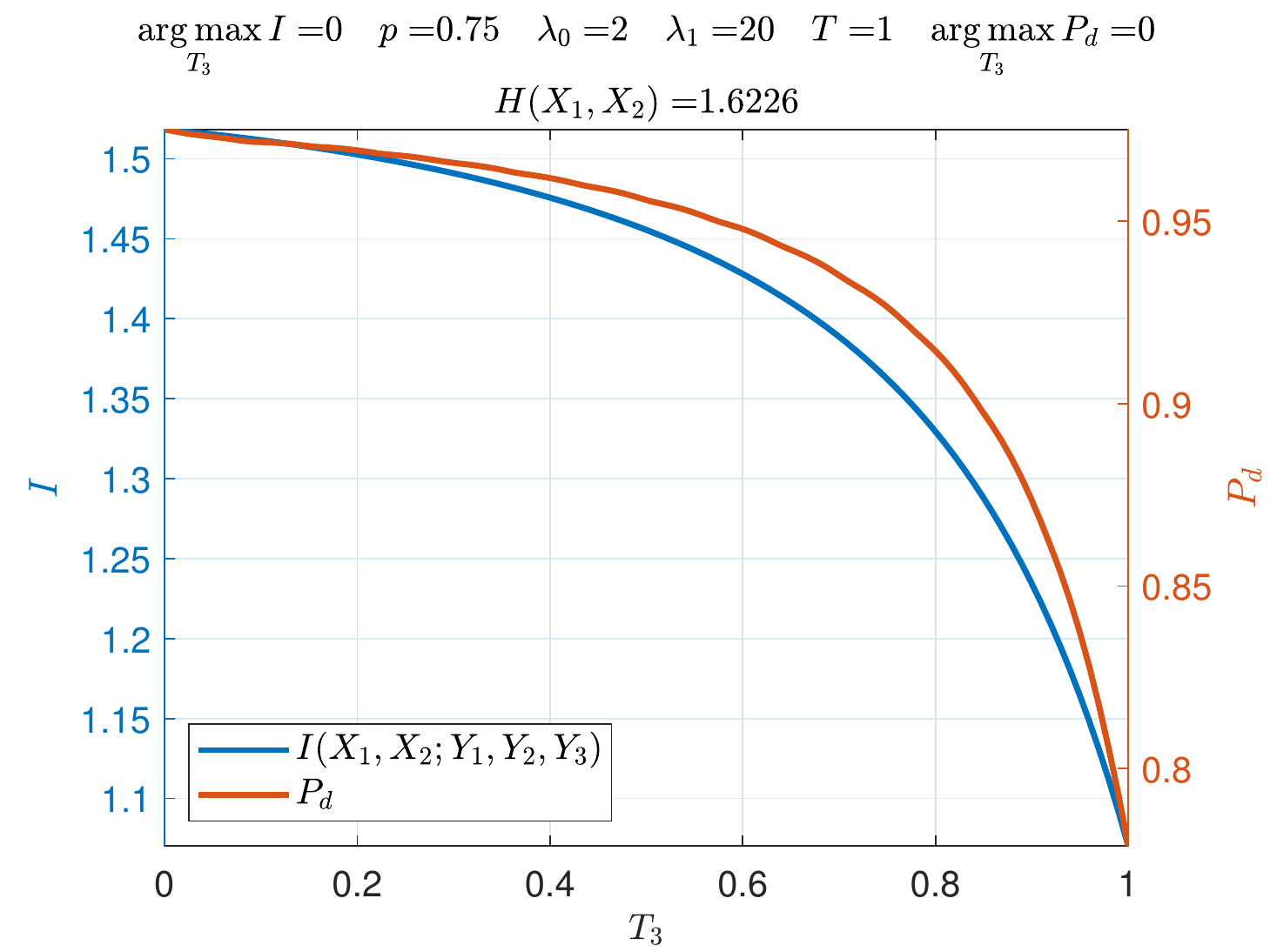}
			\caption{}
			\label{fig3c}
		\end{subfigure} %
		\begin{subfigure}[b]{.49\textwidth}
			\centering
			\includegraphics[width=1\linewidth,height=50mm]{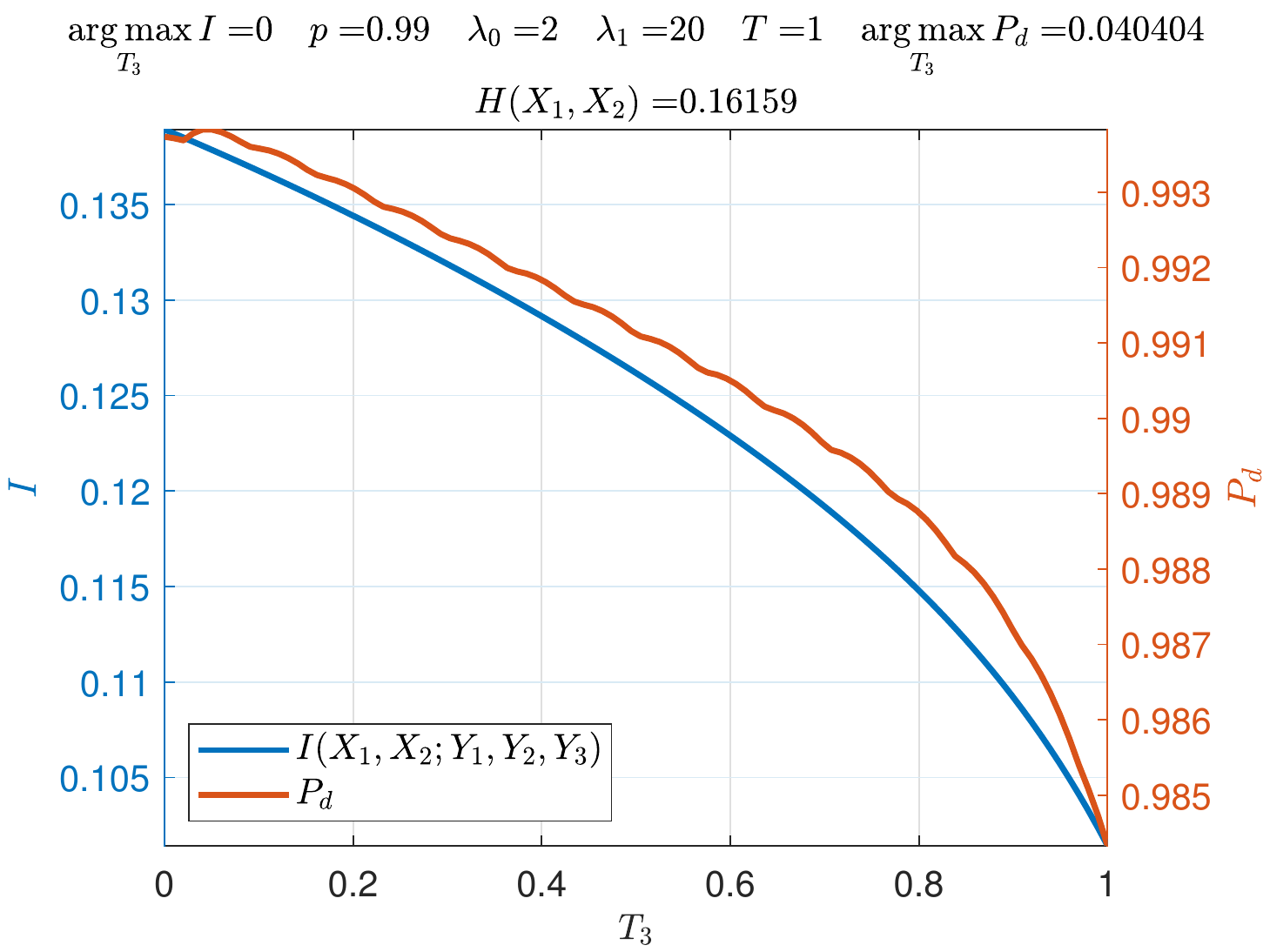}
			\caption{}
			\label{fig3d}
		\end{subfigure}
		\caption{Mutual information $ I(X;Y) $ and probability of total correct detections $P_d$ vs. $ T_3 $ for \emph{prior} probabilities of $ 0.25,0.5,0.75 $ and $ 0.99. $  \todo[disable,inline]{\texttt{\detokenize{Pd_Cd_MI_FullSupport_paper1.m}}
				\newline   	\texttt{\detokenize{MI_2.m}}                     }}
		\label{f3}
	\end{figure}
	
	\bibliographystyle{IEEEtran}
	\bibliography{ReferencesGlobal} 

\begin{thebibliography}{10}
\providecommand{\url}[1]{#1}
\csname url@samestyle\endcsname
\providecommand{\newblock}{\relax}
\providecommand{\bibinfo}[2]{#2}
\providecommand{\BIBentrySTDinterwordspacing}{\spaceskip=0pt\relax}
\providecommand{\BIBentryALTinterwordstretchfactor}{4}
\providecommand{\BIBentryALTinterwordspacing}{\spaceskip=\fontdimen2\font plus
\BIBentryALTinterwordstretchfactor\fontdimen3\font minus
  \fontdimen4\font\relax}
\providecommand{\BIBforeignlanguage}[2]{{%
\expandafter\ifx\csname l@#1\endcsname\relax
\typeout{** WARNING: IEEEtran.bst: No hyphenation pattern has been}%
\typeout{** loaded for the language `#1'. Using the pattern for}%
\typeout{** the default language instead.}%
\else
\language=\csname l@#1\endcsname
\fi
#2}}
\providecommand{\BIBdecl}{\relax}
\BIBdecl

\bibitem{hero2007foundations}
A.~O. Hero, D.~Casta{\~n}{\'o}n, D.~Cochran, and K.~Kastella,
  \emph{{Foundations and Applications of Sensor Management}}.\hskip 1em plus
  0.5em minus 0.4em\relax Springer Science \& Business Media, 2007.

\bibitem{lee2001sensor}
H.~Lee, K.~L. Teo, and A.~E. Lim, ``{Sensor scheduling in continuous time},''
  \emph{Automatica}, vol.~37, no.~12, pp. 2017--2023, 2001.

\bibitem{manyika1992information}
J.~M. Manyika and H.~F. Durrant-Whyte, ``{Information-theoretic approach to
  management in decentralized data fusion},'' in \emph{Applications in Optical
  Science and Engineering}.\hskip 1em plus 0.5em minus 0.4em\relax
  International Society for Optics and Photonics, 1992, pp. 202--213.

\bibitem{schmaedeke1993information}
W.~W. Schmaedeke, ``{Information-based sensor management},'' in \emph{Optical
  Engineering and Photonics in Aerospace Sensing}.\hskip 1em plus 0.5em minus
  0.4em\relax International Society for Optics and Photonics, 1993, pp.
  156--164.

\bibitem{yang2007mimo}
Y.~Yang and R.~S. Blum, ``{MIMO radar waveform design based on mutual
  information and minimum mean-square error estimation},'' \emph{IEEE
  Transactions on Aerospace and Electronic Systems}, vol.~43, no.~1, 2007.

\bibitem{payaro2009hessian}
M.~Payar{\'o} and D.~P. Palomar, ``{Hessian and concavity of mutual
  information, differential entropy, and entropy power in linear vector
  Gaussian channels},'' \emph{IEEE Transactions on Information Theory},
  vol.~55, no.~8, pp. 3613--3628, 2009.

\bibitem{verdu2010mismatched}
S.~Verd{\'u}, ``{Mismatched estimation and relative entropy},'' \emph{IEEE
  Transactions on Information Theory}, vol.~56, no.~8, pp. 3712--3720, 2010.

\bibitem{guo2008mutual}
D.~Guo, S.~Shamai, and S.~Verd{\'u}, ``{Mutual information and conditional mean
  estimation in Poisson channels},'' \emph{IEEE Transactions on Information
  Theory}, vol.~54, no.~5, pp. 1837--1849, 2008.

\bibitem{atar2012mutual}
R.~Atar and T.~Weissman, ``{Mutual information, relative entropy, and
  estimation in the Poisson channel},'' \emph{IEEE Transactions on Information
  Theory}, vol.~58, no.~3, pp. 1302--1318, 2012.

\bibitem{wang2014bregman}
L.~Wang, D.~E. Carlson, M.~R. Rodrigues, R.~Calderbank, and L.~Carin, ``{A
  Bregman matrix and the gradient of mutual information for vector Poisson and
  Gaussian channels},'' \emph{IEEE Transactions on Information Theory},
  vol.~60, no.~5, pp. 2611--2629, 2014.

\bibitem{boyd2004convex}
S.~Boyd and L.~Vandenberghe, \emph{{Convex Optimization}}.\hskip 1em plus 0.5em
  minus 0.4em\relax Cambridge University Press, 2004.

\bibitem{ross1996stochastic}
S.~M. Ross, \emph{{Stochastic Processes}}.\hskip 1em plus 0.5em minus
  0.4em\relax Wiley New York, 1996, vol.~2.

\bibitem{phdfahad}
\BIBentryALTinterwordspacing
M.~Fahad, ``{{Sensing Methods for Two-Target and Four-Target Detection in
  Time-Constrained Vector Poisson and Gaussian Channels}},'' Ph.D.
  dissertation, Michigan Technological University, Houghton, MI, 49931-1295,
  May 2021. [Online]. Available:
  \url{https://doi.org/10.37099/mtu.dc.etdr/1193}
\BIBentrySTDinterwordspacing

\bibitem{yeung2008information}
R.~W. Yeung, \emph{{Information Theory and Network Coding}}.\hskip 1em plus
  0.5em minus 0.4em\relax Springer Science \& Business Media, 2008.

\bibitem{cover2012elements}
T.~M. Cover and J.~A. Thomas, \emph{{Elements of Information Theory}}.\hskip
  1em plus 0.5em minus 0.4em\relax John Wiley \& Sons, 2012.

\bibitem{schonhoff2006detection}
T.~A. Schonhoff and A.~A. Giordano, \emph{{Detection and Estimation Theory and
  its Applications}}.\hskip 1em plus 0.5em minus 0.4em\relax Pearson College
  Division, 2006.

\bibitem{adell2010sharp}
J.~A. Adell, A.~Lekuona, and Y.~Yu, ``{Sharp bounds on the entropy of the
  Poisson law and related quantities},'' \emph{IEEE Transactions on Information
  Theory}, vol.~56, no.~5, pp. 2299--2306, 2010.

\bibitem{trefethen1997numerical}
L.~N. Trefethen and D.~Bau~III, \emph{{Numerical Linear Algebra}}.\hskip 1em
  plus 0.5em minus 0.4em\relax SIAM, 1997, vol.~50.

\end{thebibliography}
\end{document}